\documentclass[twocolumn]{aastex63}

\usepackage{xspace}
\usepackage{enumitem}
\usepackage[utf8]{inputenc}
\usepackage{newunicodechar,graphicx}
\usepackage{amsmath}
\usepackage{lineno}

\DeclareRobustCommand{\okina}{%
  \raisebox{\dimexpr\fontcharht\font`A-\height}{%
    \scalebox{0.8}{`}%
  }%
}
\newunicodechar{ʻ}{\okina}

\newcommand{\tess}{\textit{TESS}\xspace}
\newcommand{\ktwo}{\textit{K2}\xspace}
\newcommand{\kepler}{\textit{Kepler}\xspace}
\newcommand{\eleanor}{\textsf{eleanor}\xspace}
\newcommand{\tesscut}{\textsf{TESScut}\xspace}
\newcommand{\lightkurve}{\textsf{lightkurve}\xspace}
\newcommand{\exoplanet}{\textsf{exoplanet}\xspace}
\newcommand{\astropy}{\textsf{astropy}\xspace}
\newcommand{\astroimagej}{\textsf{astroimagej}\xspace}
\newcommand{\tapir}{\textsf{tapir}\xspace}
\newcommand{\giants}{\textsf{giants}\xspace}

\newcommand{\tstar}{\ensuremath{T_{\textrm{eff}}\:}}
\newcommand{\loggstar}{\mbox{$\log g$}\xspace}
\newcommand{\fehstar}{\mbox{$\rm{[Fe/H]}$}\xspace}
\newcommand{\vsini}{\mbox{$v\sin i$}\xspace}

\newcommand{\tic}{TIC 176956893\xspace}

\newcommand{\hoststar}{TOI-2184\xspace}

\newcommand{\starmass}{$1.53\pm0.12$\xspace} 
\newcommand{\starradius}{$2.90\pm0.14$\xspace} 
\newcommand{\teff}{$5966\pm136$\xspace} 
\newcommand{\feonh}{$0.14\pm0.08$\xspace} 
\newcommand{\age}{$2.3\pm0.8$\xspace} 
\newcommand{\starrhogcm}{$0.089\pm0.014$\xspace}
\newcommand{\logg}{$3.70\pm0.05$\xspace} 

\newcommand{\planet}{TOI-2184b\xspace}

\newcommand{\planetmass}{$0.65\pm0.16$\xspace}
\newcommand{\planetradius}{$1.017\pm0.051$\xspace}
\newcommand{\period}{$6.90683\pm0.00009$\xspace}
\newcommand{\transittime}{$1332.126\pm0.002$\xspace}
\newcommand{\K}{$53.4 \pm 8.7$\xspace}
\newcommand{\planetrho}{$0.575 \pm 0.147$\xspace}
\newcommand{\planetrhocgs}{$0.76 \pm 0.20$\xspace}
\newcommand{\ecc}{$0.08 \pm 0.07$\xspace}

\shorttitle{\tess Giants Transiting Giants I: \planet}
\shortauthors{Saunders et al.}

\graphicspath{{./}{figures/}}

\begin{document}

\title{\tess Giants Transiting Giants I: A Non-inflated Hot Jupiter Orbiting a Massive Subgiant} 

\author[0000-0003-2657-3889]{Nicholas Saunders}
\altaffiliation{NSF Graduate Research Fellow}
\affiliation{Institute for Astronomy, University of Hawaiʻi at M\=anoa, 2680 Woodlawn Drive, Honolulu, HI 96822, USA}

\author[0000-0003-4976-9980]{Samuel K.\ Grunblatt}
\altaffiliation{Kalbfleisch Fellow}
\affiliation{American Museum of Natural History, 200 Central Park West, Manhattan, NY 10024, USA}
\affiliation{Center for Computational Astrophysics, Flatiron Institute, 162 5$^\text{th}$ Avenue, Manhattan, NY 10010, USA}

\author[0000-0001-8832-4488]{Daniel Huber}
\affiliation{Institute for Astronomy, University of Hawaiʻi at M\=anoa, 2680 Woodlawn Drive, Honolulu, HI 96822, USA}

\author[0000-0001-6588-9574]{Karen A.\ Collins}
\affiliation{Center for Astrophysics \textbar \ Harvard \& Smithsonian, 60 Garden Street, Cambridge, MA 02138, USA}

\author[0000-0002-4625-7333]{Eric L.\ N.\ Jensen}
\affiliation{Department of Physics \& Astronomy, Swarthmore College, Swarthmore PA 19081, USA}

\author[0000-0001-7246-5438]{Andrew Vanderburg}
\affiliation{Center for Astrophysics \textbar \ Harvard \& Smithsonian, 60 Garden Street, Cambridge, MA 02138, USA}
\affiliation{Department of Astronomy, The University of Texas at Austin, Austin, TX 78712, USA}

\author[0000-0002-9158-7315]{Rafael Brahm}
\affiliation{Facultad de Ingeniería y Ciencias, Universidad Adolfo Ibáñez, Av. Diagonal las Torres 2640, Peñalolén, Santiago, Chile}
\affiliation{Millennium Institute for Astrophysics, Chile}

\author[0000-0002-5389-3944]{Andrés Jordán}
\affiliation{Facultad de Ingeniería y Ciencias, Universidad Adolfo Ibáñez, Av. Diagonal las Torres 2640, Peñalolén, Santiago, Chile}
\affiliation{Millennium Institute for Astrophysics, Chile}

\author[0000-0001-9513-1449]{Néstor Espinoza}
\affiliation{Space Telescope Science Institute, 3700 San Martin Drive, Baltimore, MD 21218, USA}

\author[0000-0002-1493-300X]{Thomas Henning}
\affiliation{Max-Planck-Institut für Astronomie, Königstuhl 17, 69117 Heidelberg, Germany}


\author[0000-0002-5945-7975]{Melissa J. Hobson}
\affiliation{Millennium Institute for Astrophysics, Chile}
\affiliation{Instituto de Astrofísica, Facultad de Física, Pontificia Universidad Católica de Chile, Av. Vicuña Mackenna 4860, 782-0436 Macul, Santiago, Chile}

\author[0000-0002-8964-8377]{Samuel N. Quinn}
\affiliation{Center for Astrophysics \textbar \ Harvard \& Smithsonian, 60 Garden Street, Cambridge, MA 02138, USA}

\author[0000-0002-4891-3517]{George Zhou}
\affiliation{University of Southern Queensland, Centre for Astrophysics, West Street, Toowoomba, QLD 4350 Australia}

\author[0000-0003-1305-3761]{R. Paul Butler}
\affiliation{Carnegie Institution for Science, Earth \& Planets Laboratory, 5241 Broad Branch Road NW, Washington DC 20015, USA}

\author{Lisa Crause}
\affiliation{South African Astronomical Observatory, P.O. Box 9, Observatory 7935, Cape Town, South Africa}

\author[0000-0002-4236-9020]{Rudi B. Kuhn}
\affiliation{South African Astronomical Observatory, P.O. Box 9, Observatory 7935, Cape Town, South Africa}

\author[0000-0002-5136-7983]{K. Moses Mogotsi}
\affiliation{South African Astronomical Observatory, P.O. Box 9, Observatory 7935, Cape Town, South Africa}

\author{Coel Hellier}
\affiliation{Astrophysics Group, Keele University, Staffordshire ST5 5BG, UK}

\author[0000-0003-4540-5661]{Ruth Angus}
\affiliation{American Museum of Natural History, 200 Central Park West, Manhattan, NY 10024, USA}
\affiliation{Center for Computational Astrophysics, Flatiron Institute, 162 5$^\text{th}$ Avenue, Manhattan, NY 10010, USA}
\affiliation{Department of Astronomy, Columbia University, 550 West 120$^\text{th}$ Street, New York, NY, USA}

\author[0000-0002-0842-863X]{Soichiro Hattori}
\affiliation{American Museum of Natural History, 200 Central Park West, Manhattan, NY 10024, USA}
\affiliation{New York University Abu Dhabi, Abu Dhabi, United Arab Emirates}

\author[0000-0003-1125-2564]{Ashley Chontos}
\altaffiliation{NSF Graduate Research Fellow}
\affiliation{Institute for Astronomy, University of Hawaiʻi at M\=anoa, 2680 Woodlawn Drive, Honolulu, HI 96822, USA}


\author{George R.\ Ricker}
\affiliation{Department of Physics, and Kavli Institute for Astrophysics and Space Research, Massachusetts Institute of Technology, 77 Massachusetts Ave., Cambridge, MA 02139, USA}

\author[0000-0002-4715-9460]{Jon M.\ Jenkins}
\affiliation{NASA Ames Research Center, Moffett Field, CA, 94035}

\author[0000-0002-1949-4720]{Peter Tenenbaum}
\affiliation{NASA Ames Research Center, Moffett Field, CA, 94035}
\affiliation{SETI Institute, 189 Bernardo Ave, Suite 200 Mountain View, CA 94043, USA}

\author[0000-0001-9911-7388]{David W. Latham}
\affiliation{Center for Astrophysics \textbar \ Harvard \& Smithsonian, 60 Garden Street, Cambridge, MA 02138, USA}

\author[0000-0002-6892-6948]{Sara Seager}
\affiliation{Department of Physics, and Kavli Institute for Astrophysics and Space Research, Massachusetts Institute of Technology, 77 Massachusetts Ave., Cambridge, MA 02139, USA}
\affiliation{Department of Earth, Atmospheric, and Planetary Sciences, Massachusetts Institute of Technology, 77 Massachusetts Ave., Cambridge, MA 02139, USA}
\affiliation{Department of Aeronautics and Astronautics, Massachusetts Institute of Technology, 77 Massachusetts Ave., Cambridge, MA 02139, USA}

\author{Roland K.\ Vanderspek}
\affiliation{Department of Physics, and Kavli Institute for Astrophysics and Space Research, Massachusetts Institute of Technology, 77 Massachusetts Ave., Cambridge, MA 02139, USA}

\author[0000-0002-4265-047X]{Joshua N.\ Winn}
\affiliation{Department of Astrophysical Sciences, Princeton University, 4 Ivy Lane, Princeton, NJ 08544, USA}


\author[0000-0003-2163-1437]{Chris Stockdale}
\affiliation{Hazelwood Observatory, Australia}

\author[0000-0001-5383-9393]{Ryan Cloutier}
\altaffiliation{Banting Fellow}
\affiliation{Center for Astrophysics \textbar \ Harvard \& Smithsonian, 60 Garden Street, Cambridge, MA 02138, USA}


\begin{abstract}

    While the population of confirmed exoplanets continues to grow, the sample of confirmed transiting planets around evolved stars is still limited. We present the discovery and confirmation of a hot Jupiter orbiting \hoststar (\tic), a massive evolved subgiant ($M_\star=$ \starmass $M_\odot$, $R_\star=$ \starradius $R_\odot$) in the \tess Southern Continuous Viewing Zone. The planet was flagged as a false positive by the \tess Quick-Look Pipeline due to periodic systematics introducing a spurious depth difference between even and odd transits. Using a new pipeline to remove background scattered light in \tess Full Frame Image (FFI) data, we combine space-based \tess photometry, ground-based photometry, and ground-based radial velocity measurements to report a planet radius of $R_p=$ \planetradius $R_J$ and mass of $M_p=$ \planetmass $M_J$. For a planet so close to its star, the mass and radius of \planet are unusually well matched to those of Jupiter. We find that the radius of \planet is smaller than theoretically predicted based on its mass and incident flux, providing a valuable new constraint on the timescale of post-main-sequence planet inflation. The discovery of \planet demonstrates the feasibility of detecting planets around faint (\tess magnitude $>12$) post-main sequence stars and suggests that many more similar systems are waiting to be detected in the \tess FFIs, whose confirmation may elucidate the final stages of planetary system evolution.
    
    


\end{abstract}

\section{Introduction} \label{sec:intro}

The Transiting Exoplanet Survey Satellite (\tess; \citealt{ricker2014}) has observed over 80\% of the sky, enabling the discovery of a predicted $\sim$14,000 planets \citep{sullivan2015, barclay2018}. The space telescope observes most of its targets in the Full Frame Images (FFIs) with a 30-minute observing cadence, and has completed a full year of observations in both the northern and southern hemispheres. Each hemisphere was split into 13 sectors that stretched from the ecliptic pole to the ecliptic plane, which were observed for $\sim$27 days at a time. Targets near the ecliptic pole appear in all sectors, allowing for a full year of photometry in what is known as the Continuous Viewing Zone (CVZ), while targets closer to the ecliptic plane were observed in fewer sectors. According to the NASA Exoplanet Science Institute (NExScI) archive\footnote{Data retrieved from \href{https://nexsci.caltech.edu}{nexsci.caltech.edu} March 17, 2021.}, \tess has already discovered $>100$ confirmed planets and $>2,500$ candidates. Of the 120 confirmed planets, only a handful orbit evolved host stars; among those is TOI-197.01b, the first \tess planet discovery orbiting an evolved host with an asteroseismic detection \citep{huber2019}. There have also been detections of planets orbiting subgiant stars for which asteroseismic detections were not possible, for instance TOI-813b (also known as Planet Hunters \tess I; \citealt{eisner2020}), a Saturn-sized planet orbiting a subgiant.


Planets orbiting evolved stars are a poorly understood population. For example, the source of anomalously large hot Jupiters with radii up to 2 $R_J$ on short-period orbits around evolved stars has been debated for over twenty years (\citealt{guillot1996, burrows2000, batygin10, grunblatt2017}). The leading theories for this planet inflation can be separated into two classes—in class I theories, a planet begins its life inflated, cools and contracts during the majority of its main sequence lifetime, but then begins to re-inflate as its host star evolves off the main sequence into a red giant \citep{lopez2016}. In class II theories, the planet similarly cools and contracts, but this cooling is delayed, resulting in a planet which appears equally inflated during main sequence and post-main sequence phases. The deviation between the model predictions is most pronounced in the post-main-sequence phase of stellar evolution, and a larger sample of short-period gas giants at various stages of host star evolution will help settle this debate. 

Planet detections around evolved stars also provide constraints on planet inspiral and engulfment due to tidal dissipation, the timescales of which are still poorly understood \citep{villaver2009, grunblatt2018}. Tidal orbital decay is the transfer of angular momentum from a planet’s orbit to its host star’s rotation, causing the planet to spiral into the star \citep{zahn1977, hut1981}. The timescale of orbital decay depends on the tidal quality factor $Q'_\star$ . By measuring the deviation from a constant orbital period over the course of many observations for a hot Jupiter around subgiant stars, the inspiral timescale and value of $Q'_\star$ for the host can be constrained (e.g. \citealt{levrard2009}; \citealt{chontos2019}). Planets around evolved stars can also contribute to the resolution of debates about the dependence of planet occurrence on stellar mass, a hotly debated topic over the last decade \citep{johnson2010, lloyd2013, schlaufman13, ghezzi2015}. 

The majority of short-cadence observing slots in \tess and similar surveys are reserved for main sequence stars, in part due to the difficulty of detecting planets around luminous and noisy subgiants and Red Giant Branch (RGB) stars. This makes the FFIs an ideal dataset to find new planets orbiting evolved stars. There are currently \tess teams identifying candidates in the \tess data---NASA runs the pipeline at the Science Processing Operations Center (SPOC; \citealt{jenkins2016}) at Ames Research Center, and the \tess Science Office at MIT manages the Quick-Look Pipeline (QLP; \citealt{huang2020}). As with many planet-search efforts, these pipelines are optimized to detect small planets around solar type stars. To accomplish this, main sequence stars are given priority, as are nearer, brighter targets. This means there is a large population of evolved potential host stars that remains to be explored in detail. This is particularly true for the more distant and therefore fainter subgiants and giants. 

In this paper we present the first results from a search for planets around evolved stars in \tess, including the development of a novel pipeline to remove background scattered light from \tess FFIs with an emphasis on evolved and faint stars. We also confirm and characterize \planet, a hot Jupiter orbiting a subgiant star, which was initially labeled as a false positive by QLP pipeline due to a depth difference between even and odd transits caused by \tess background scattered light, and was not searched further until our identification of the system as a potential planet candidate. This timeline demonstrates the difficulty of large scale planet-search efforts and the benefit to a focused search for planets specifically orbiting evolved stars. Our fit to the data indicates that \planet is not significantly inflated, and is among the smallest hot Jupiters of similar mass, providing new constraints on timescales and mechanisms for planet inflation in evolved systems.

\section{\tess Photometry} \label{sec:photo}

\subsection{Target Selection}

We selected evolved stars with the explicit motivation of identifying new planets orbiting subgiants and RGB stars. Using the \tess Input Catalog (TIC), we made cuts based on color, magnitude, and Gaia parallax in order to limit our sample to evolved stars. The stars in our targeted sample were selected with temperatures between 4500 and 5500 K, surface gravities of 2.9 $<$ log(\textit{g}) $<$ 3.5, and \tess magnitude T $<$ 13. These are adapted from the cuts made to the \ktwo sample in \cite{grunblatt2019}. 

\subsection{Light Curve Generation and Background Correction} \label{sec:systematics}

A dominant source of systematic trends in the \tess FFI light curves is due to sunlight reflected off of the Earth and Moon, indirectly illuminating the detector. Due to the periodic orbit of \tess, this incident light varies dramatically on relatively short timescales. The background reflected light variation can be orders of magnitude higher than the low signal-to-noise ratio (SNR) transits in the underlying light curve. It is therefore necessary to create a model for the systematic trends which can be subtracted to isolate the desired signal and make transit detection feasible. 

To begin, we used the \tesscut tool \citep{brasseur2019} to download an 11x11 pixel cutout around each target from the Mikulski Archive for Space Telescopes (MAST). The cutout is made from the SPOC-generated FFI data cubes hosted on MAST, and we perform our own simple-aperture photometry to generate an uncorrected light curve. We then created an aperture mask for each target by taking a contiguous set of pixels connected to the central pixel which are $3\sigma$ above the median flux in the cutout. To begin our correction for contributions from scattered Earth light, we first created a design matrix from the flux light curves of each background pixel outside the target aperture mask. 
%
%

We then performed principal component analysis (PCA) on these background pixel data to find a set of basis vectors for our background flux model to create a design matrix, $X$:
\[
X = 
\begin{bmatrix}
p_{11} & p_{12} & \dots & p_{1j}\\
p_{21} & p_{22} & \dots & p_{2j} \\
\vdots & \vdots & \ddots & \vdots \\
p_{i1} & p_{i2} & \dots & p_{ij} \\
\end{bmatrix}
\]
where $p_{ij}$ is the principal component $j$ of the background pixel light curves at time $i$. Assuming the scattered light background incident on the \tess detector can be modeled as some combination of the signals in each of the background pixels, we can create a scattered light model $m$ from $X$ by placing a coefficient $w_j$ on each regressor column vector $p_{j}$. This allows us to define the model as a linear combination:
\[
m = X \cdot w.
\]
Using the \textsf{RegressionCorrector} framework in the \textsf{lightkurve} Python package \citep{lightkurve}, we fit coefficients to the column vectors of the design matrix $X$ to create a scattered light model. To optimize the coefficient fit, we minimized the square difference between the model and data, represented by $\chi^2$
\[
\chi^2 = \sum_i \frac{(f_i-m_i)^2}{\sigma_i^2}
\]
where $f_i$ is the simple aperture photometry (SAP) flux value at time $i$ and $\sigma_i$ is the flux uncertainty at time $i$. We want to find the values of $w$ which minimize $\chi^2$, which we accomplish by solving
\[
\frac{\partial \chi^2}{\partial w}=0.
\]
We also want to consider the covariance between points in the SAP light curve to account for stellar variability, so we replace $\sigma$ with a matrix $\Sigma$ which includes the uncertainties $\sigma$ along the diagonal and flexible priors for covariance on the off diagonal to prevent overfitting. 
Because $X$ and $\Sigma$ are matrices, and $y$ and $w$ are arrays, this becomes a generalized least squares problem. We can solve for $w$:
\[
w = (X^\top\cdot \Sigma^{-1}\cdot X)^{-1}\cdot (X^\top\cdot\Sigma^{-1}\cdot f)
\]
which is used to compute the corrected light curve $y$
\[
y = f-X\cdot w.
\]
The optimized scattered light model was subtracted from the raw flux light curve to produce a background-corrected light curve. This procedure is similar to the Pixel Level Decorrelation method applied to the \textit{Spitzer} Space Telescope by \cite{deming2015} and the \textit{K2} mission by \cite{luger2016,luger2018}. The major difference between these applications is that uncorrected \tess observations are dominated by background scattered light while \textit{Spitzer} and \textit{K2} primarily suffer from instrumental signal introduced by spacecraft motion during observations. To account for this, our approach focuses on choosing regressors exclusively from background pixels to ensure our systematics model captures this high-amplitude signal.

\begin{figure}[h]
    \centering
    \includegraphics[width=0.45\textwidth]{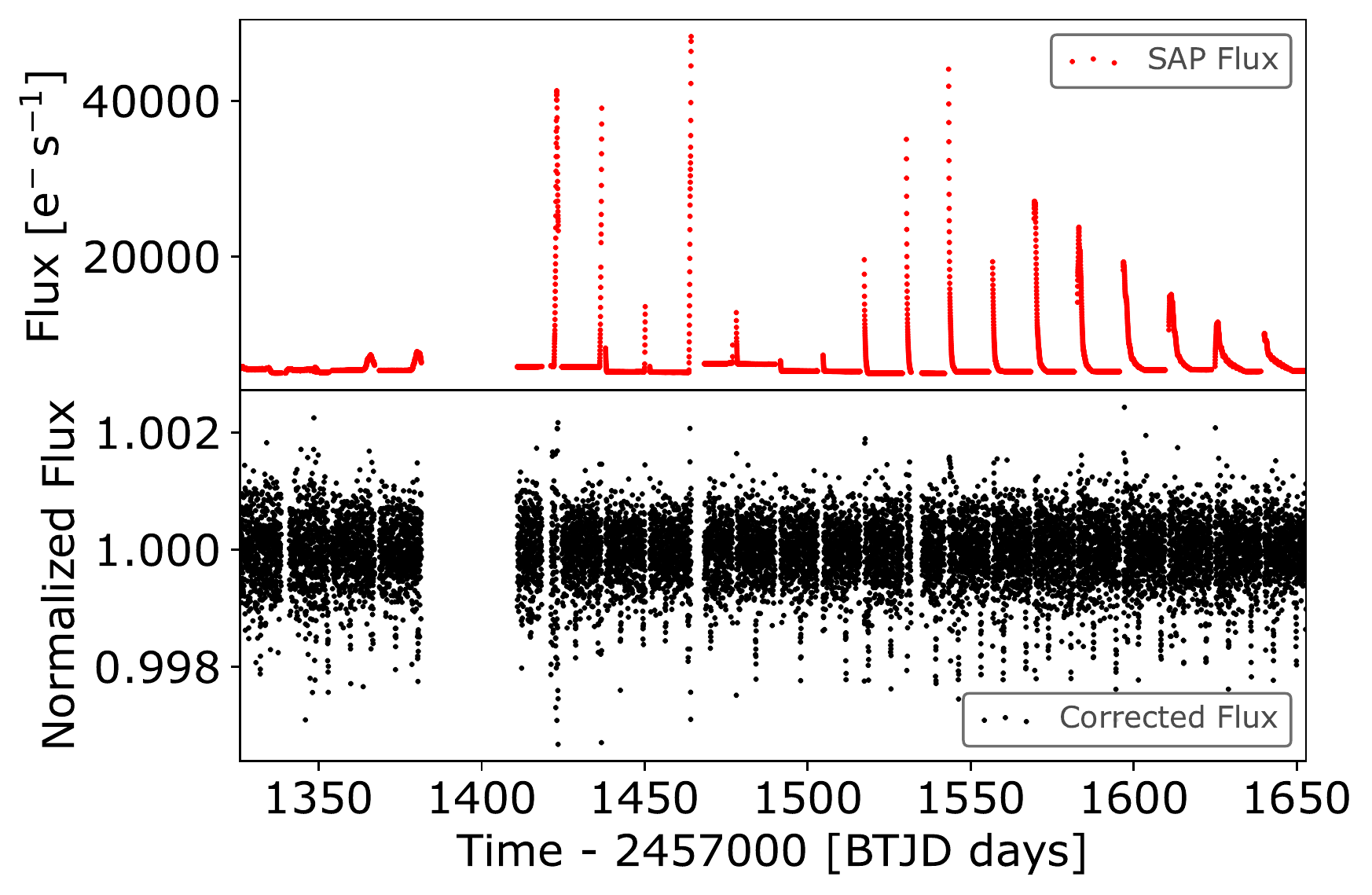}
    \caption{The light curve for \planet before (top panel) and after (bottom panel) applying our de-trending method.}
    \label{fig:noise}
\end{figure}

To clean the light curve more thoroughly, we masked the transits of \planet, then identified and removed data points that were greater than or less than the median flux by at least $6\sigma$ (for the standard deviation of the flux light curve $\sigma$). Additionally, we applied a Gaussian filter to smooth trends on timescales greater than $\sim2$ days. A comparison of the uncorrected SAP flux light curve and the final corrected light curve for \planet can be found in Figure \ref{fig:noise}.

\subsection{Transit Search Pipeline} \label{sec:search}

We incorporated the algorithm described in the previous section into the \giants\footnote{\href{https://github.com/nksaunders/giants}{https://github.com/nksaunders/giants}} Python package for accessing, de-trending, and searching \tess observations for periodic transit signals, with an emphasis on detecting planets around subgiant and RGB stars. To search for transits, we used the \textsf{astropy.timeseries} implementation of the box least squares (BLS) method \citep{kovacs2002}. 

The \giants pipeline produces a one-page PDF summary for each target including the following vetting materials:
\begin{itemize}
    \item The full de-trended light curve (de-trending methods described in detail in \S \ref{sec:systematics})
    \item Lomb-Scargle periodgram \citep{lomb1976, scargle1982} to identify stellar oscillations in the flux light curve
    \item Box least squares (BLS) periodogram \citep{kovacs2002} to vet the quality of the most likely potential transit in each light curve
    \item Flattened light curve folded with the period of maximum power in the BLS periodogram
    \item Folded light curves of exclusively even and odd transits to identify the existence of a depth difference (see \S \ref{sec:evenodd})
    \item Initial transit fit using the \textsf{ktransit} Python package \citep{barclay2015}.
\end{itemize}

The vetting sheet for \planet can be found in Figure \ref{fig:vetting} in the Appendix. Together, these elements provide the necessary information for transiting planet candidate vetting. After identifying the transit signal in the \giants light curve, we independently verified the presence of this transit in the \tess photometry by generating a \tess light curve for \hoststar using the \eleanor pipeline \citep{feinstein2019} and QLP pipeline \citep{huang2020}. We applied the same outlier rejection and Gaussian smoothing as described in \S \ref{sec:systematics} and performed our transit search on the \eleanor and QLP light curves. When we applied the same BLS search, we identified an eclipse signal whose period agreed with that found in our corrected light curve within errors in light curves from each of the other pipelines.

\subsection{Odd/Even Transit Depths} \label{sec:evenodd}

In the case of \planet, after removing the background scattered light signal, additional \tess systematics presented a new challenge. The orbit of \tess has a period of 13.7 days, which is comparable to the orbital period of many transiting exoplanets, particularly hot and warm Jupiters. This orbit produces a periodic scattered light signal in \tess observations due to scattered light from the Earth and thermal sensitivity changes of the detector during data downlinks \citep{luger2019}.

We measure the orbital period of \planet to be \period days, which is roughly half the orbital period of \tess. Additionally, for the majority of the year \tess spent observing the southern CVZ, every other transit of \planet occurred near or during a data downlink. This caused more transit dilution in every other transit, leading to a slight difference in measured depth in odd and even transits. A difference in depth of alternating eclipses is a characteristic signal of eclipsing binaries (EBs), which display alternating deep primary eclipses and shallower secondary eclipses. For this reason, the slight difference in even/odd transit depth mimics the signal of a background EB. This caused the system to be rejected as a TOI by the QLP pipeline, and delayed it from being studied with other pipelines, such as SPOC, until after we had flagged the target as a community TOI (CTOI). After our identification of \planet as a CTOI, it was vetted by the TESS Follow-up Observing Program (TFOP)\footnote{\href{https://tess.mit.edu/followup}{https://tess.mit.edu/followup}} and upgraded to TOI status. Targets with TOI status were observed with 2-minute cadence during the \tess extended mission when the field was revisited in Year 3. While we do not include the extended mission data in our fit, we utilize the SPOC Data Validation reports to confirm that our results are consistent with the updated observations and discuss this extended mission data validation at the end of this section.

\begin{figure*}[ht!]
    \centering
    \includegraphics[width=0.9\textwidth]{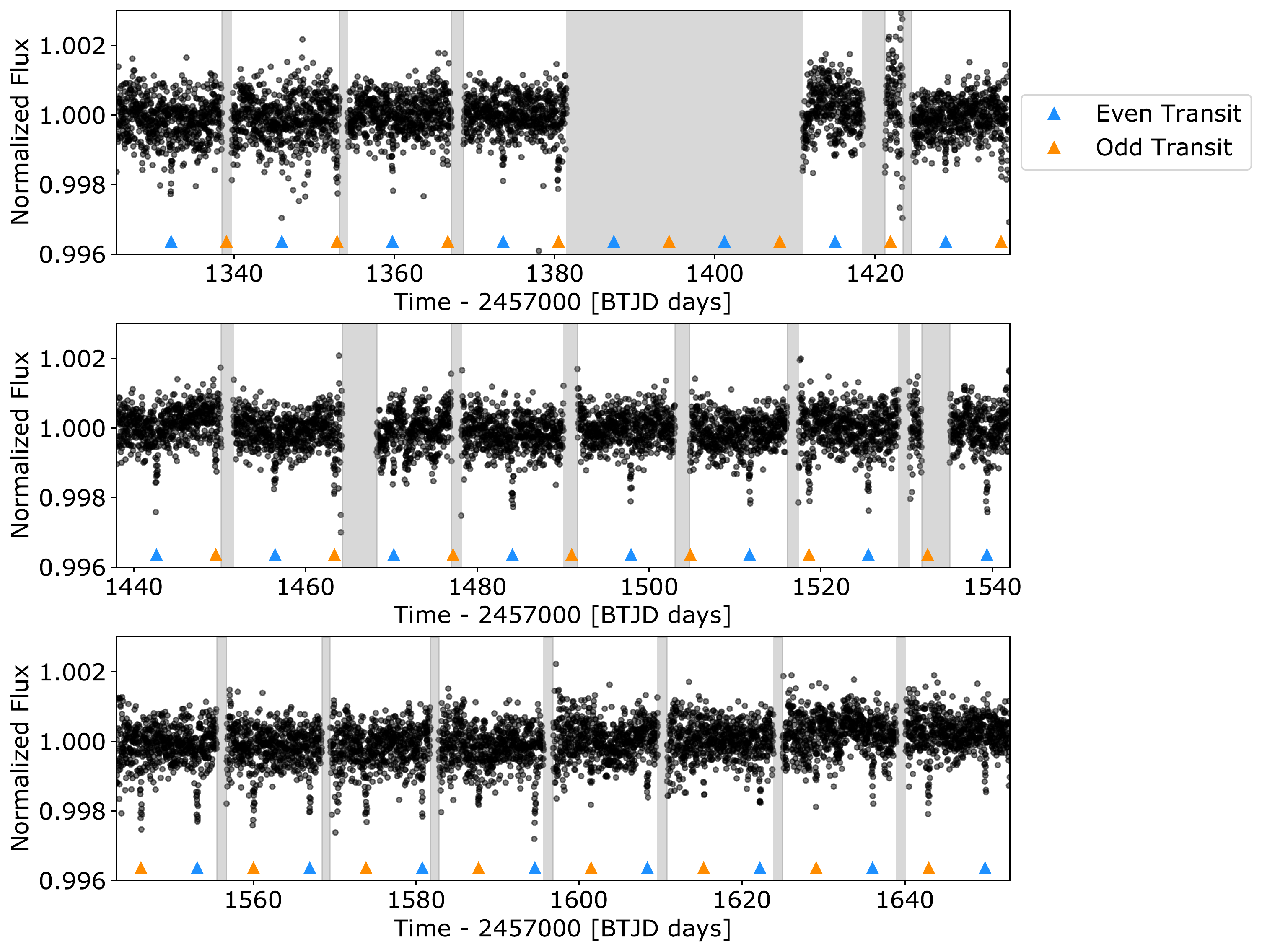}
    \caption{The full systematics-corrected flux light curve for \planet with transit times marked by triangles. The alternating triangle colors represent the even and odd transits. The gray sections mark the data gaps due to data downlinks or the spacecraft entering safe mode. While the even transits (blue) fall almost exclusively in the middle of continuous data collection, the odd transits (orange) fall primarily near the beginning or end of a data gap when the \tess detector is experiencing increased scatter due to thermal sensitivity variation and scattered light. This trend is most present in the first seven observing sectors (top two rows), and disappears for the final four sectors (bottom row).}
    \label{fig:lcevenodd}
\end{figure*}

Figure \ref{fig:lcevenodd} shows the flux light curve for the entire year in the CVZ, with transit times marked by colored triangles. Even and odd transits are differentiated by alternating colors, and it is apparent that odd transits (marked by orange triangles) fall near data downlinks more frequently than even transits (marked by blue triangles), particularly during the first 200 days of observations. This effect increases the scatter of the in-transit light curve and causes a slight difference in the measured depth due to the loss of precision.

\begin{figure}[ht!]
    \centering
    \includegraphics[width=.45\textwidth]{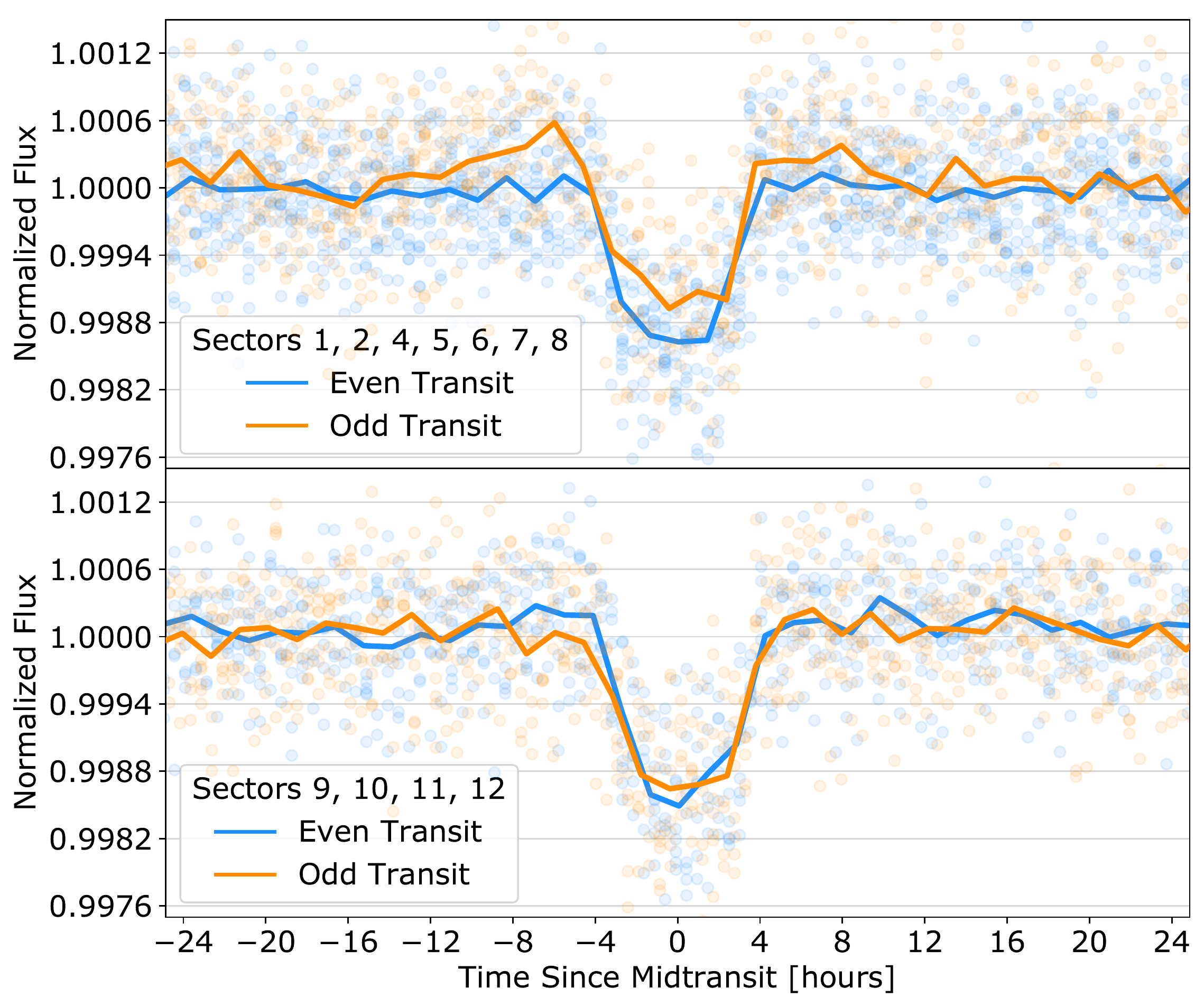}
    \caption{A comparison between the even and odd transits of \planet for the first seven sectors (top) and the final four sectors (bottom). Faint points show the original phase-folded photometry and solid lines show the binned photometry.}
    \label{fig:evenodd_compare}
\end{figure}

Examining the timing of alternating transits reveals that during the final four sectors in which \hoststar was observed, both odd and even transits occurred during observations and further from data downlinks. To ensure that this odd/even difference is a strictly systematic effect introduced by the unlucky timing of transits, we analyzed the phase-folded, alternating even and odd transits for two distinct observing periods---first: sectors 1, 2, 4, 5, 6, 7, and 8, and second: sectors 9, 10, 11, and 12. These two groups of sectors correspond to the top two panels of Figure \ref{fig:lcevenodd} and the bottom panel of Figure \ref{fig:lcevenodd}, respectively. If this were a solely de-trending systematic effect, the odd/even depth difference would disappear in the final set of sectors. The comparison between these can be found in Figure \ref{fig:evenodd_compare}.

The odd transits, which fell preferentially close to the data downlinks for the first seven sectors during which \hoststar was observed by \tess, demonstrate an unusual systematic offset in the folded light curve, showing a trend that rises prior to ingress and presents a shallower transit depth than its even counterparts. This can be compared to the final four sectors during which the target was observed, which shows no meaningful difference in transit depth. This leads us to conclude that the odd/even difference that caused this candidate to be rejected from previous pipelines was purely due to systematic \tess trends that resulted from poor data quality and transit dilution near data gap.

While these circumstances presented a particularly tricky challenge to the detection of \planet, we note that the highly periodic \tess orbit makes detection of planets with periods that are fractions or multiples of the spacecraft orbit difficult in general. However, the extended baseline of the TESS extended mission improves the phase coverage of TESS observations and provides additional evidence to support the confirmation of \planet.

After the \tess prime mission, \planet was placed on the 2-minute candidate target list for the first year of the \tess extended mission and was observed in sectors 27, 28, 29, 31, 32, 34, 35, and 36. The SPOC pipeline conducted a transit search of these data \citep{jenkins2002,jenkins10}, detecting the signature of TOI-2184b and generating Data Validation reports for each sector. The transit signature passed all of the data validation tests, including the odd/even transit depth test \citep{twicken2018} and was fitted with a limb-darkened transit model \citep{li2019}. The Data Validation reports do not show strong evidence of odd/even transit depth differences, which may be due to the fact that the orbital phase has shifted sufficiently in the interim such that the instrumental systematics that drove the asymmetric results for the QLP pipeline in Year 1 observations are weaker in Year 3. 

\section{Ground-based Follow-up} \label{sec:ground}

\subsection{Transit Observation}

The TESS pixel scale is $\sim 21\arcsec$ pixel$^{-1}$, and photometric apertures typically extend out to roughly 1 arcminute, which generally results in multiple stars blending in the TESS aperture. An eclipsing binary in one of the nearby blended stars could mimic a transit-like event in the large TESS aperture. We conducted ground-based photometric follow-up observations as part of TFOP with much higher spatial resolution to confirm that the transit signal is occurring on-target, or on a star so close to TOI-2184 that it was not detected by Gaia DR2. The ground-based observations also confirm or refine the TESS ephemeris, transit depth, and transit duration.

We observed a predicted ingress and a predicted egress of \planet on UTC 2020 February 12 in Pan-STARRS $z$-short band from the Las Cumbres Observatory Global Telescope \citep[LCOGT;][]{Brown:2013} 1.0\,m nodes at South Africa Astronomical Observatory (SAAO) and Cerro Tololo Inter-American Observatory (CTIO), respectively. We used the {\sf TESS Transit Finder}, which is a customized version of the {\sf Tapir} software package \citep{Jensen:2013}, to schedule our transit observations. The $4096\times4096$ LCOGT SINISTRO cameras have an image scale of $0\farcs389$ per pixel, resulting in a $26\arcmin\times26\arcmin$ field of view. The images were calibrated by the standard LCOGT {\sf BANZAI} pipeline \citep{McCully:2018}, and photometric data were extracted with {\sf AstroImageJ} \citep{Collins:2017}. The images were focused and have typical stellar point-spread-functions with a full-width-half-maximum (FWHM) of $\sim 2\arcsec$, and circular apertures with radius $5\arcsec$ were used to extract the differential photometry. The photometric apertures exclude most of the flux from the nearest Gaia DR2 star (TIC 765203880), which is $7\farcs2$ southwest of TOI-2184, so we conclude that the TESS-detected transit signal is on-target relative to known Gaia DR2 stars. 

\begin{figure}[ht!]
    \centering
    \includegraphics[width=0.5\textwidth]{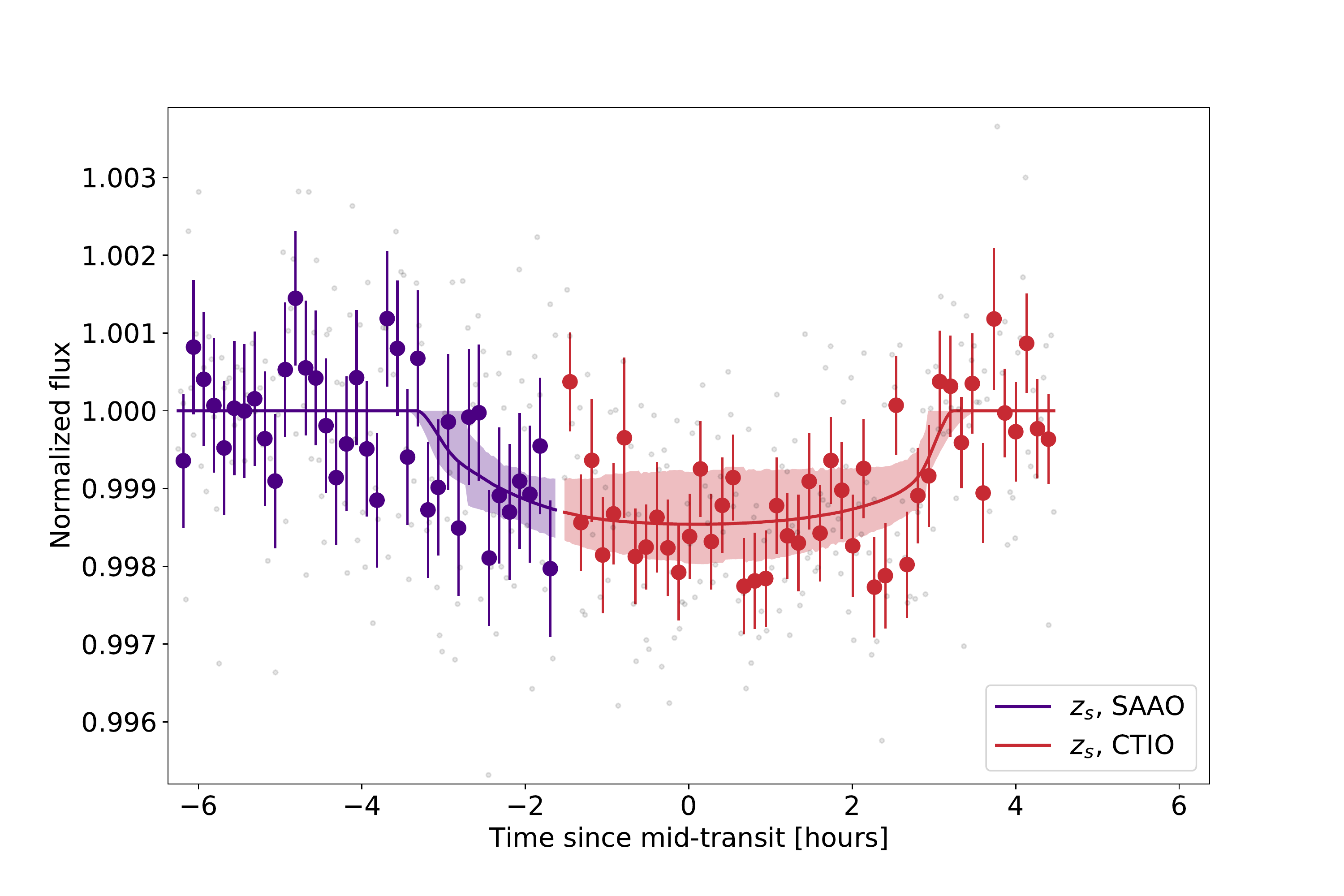}
    \caption{The LCOGT light curve and model for \planet in Pan-STARRS $z$-short band. The light gray symbols show the unbinned photometry. The purple and red symbols show the photometry from SAAO and CTIO, respectively, in 8-minute bins. The transit model is shown as purple and red lines, and the shaded regions represent the 68.3\% highest-density region in the posteriors. The fitted parameters are consistent with the transit parameters from the TESS data within $1\sigma$ uncertainty (see Sections \ref{sec:ground} and \ref{sec:planet}).}
    \label{fig:LCOGT_lightcurve}
\end{figure}

An independent fit to the ground-based data finds ${\rm R_p}=0.035^{+0.015}_{-0.013}~{\rm R_{star}}$, ${\rm duration}=6.5^{+0.6}_{-0.5}~{\rm hours}$, and an orbital period ${\rm P}=6.906895^{+0.000212}_{-0.000094}~{\rm days}$ (using the reference epoch we derive from the TESS data), and are all within $1\sigma$ of values extracted from our simultaneous fit to the TESS and RV data (see Section \ref{sec:planet}). The LCOGT light curve and independent model fit are presented in Figure \ref{fig:LCOGT_lightcurve}. The follow-up light curve data are available at ExoFOP-TESS\footnote{\href{https://exofop.ipac.caltech.edu/tess}{https://exofop.ipac.caltech.edu/tess}}. 

\subsection{High-Resolution Imaging}

\begin{figure}[ht!]
    \centering
    \includegraphics[width=0.45\textwidth]{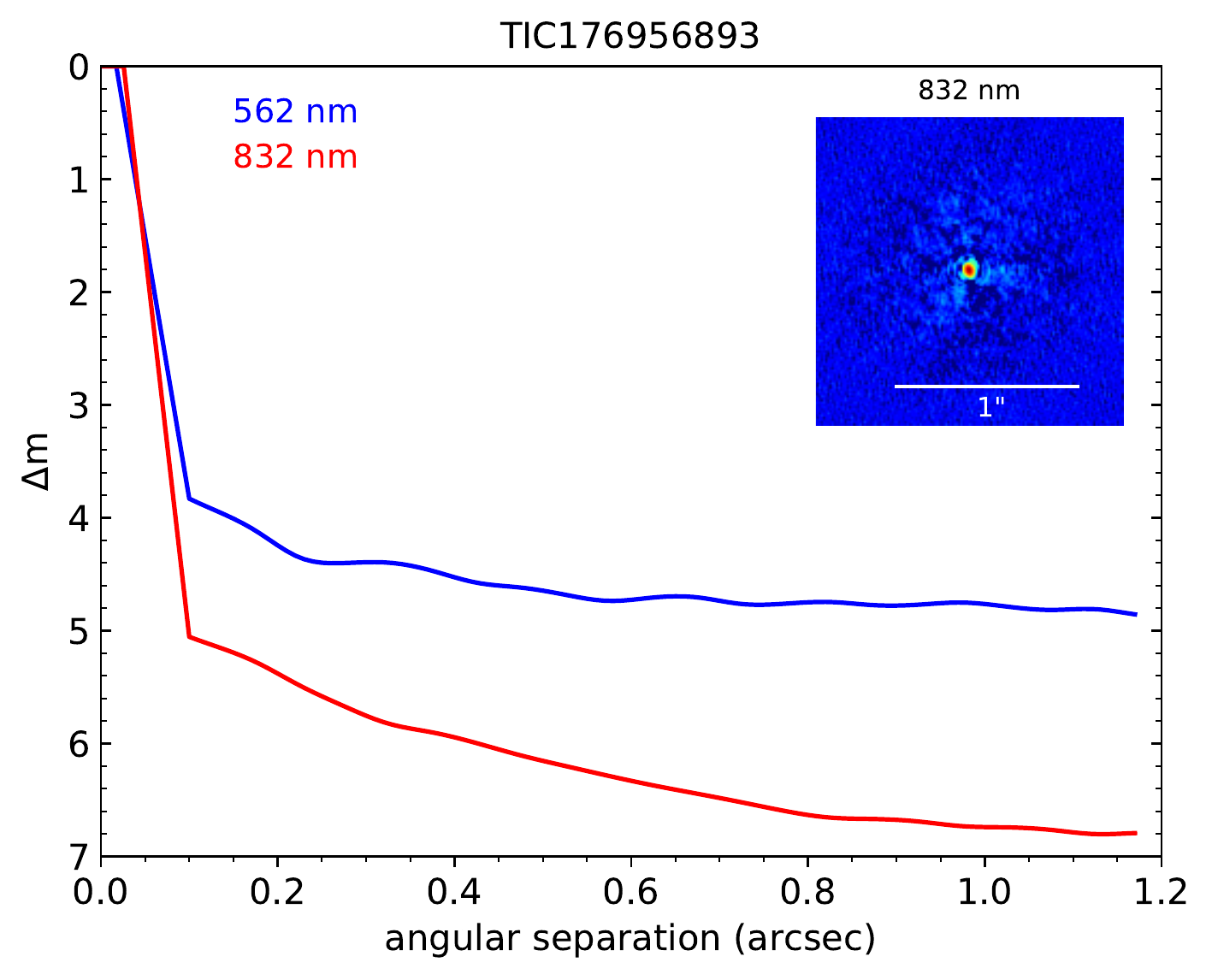}
    \caption{The contrast curve for \hoststar in two bands---562 nm and 832 nm. These observations were taken on March 13, 2020 using the Gemini South telescope at CTIO. There are no significant spikes in the contrast curve above $\Delta m > 4$ in the 562 nm observation and $\Delta m > 5$ in the 832 nm observation, indicating that \hoststar has no close companions. The inset figure is a high contrast speckle image centered on \hoststar.}
    \label{fig:contrast}
\end{figure}

In order to search for close stellar companions to \hoststar, we also obtained a high-resolution speckle image of \hoststar with the speckle interferometric instrument on the Gemini South telescope at CTIO \citep{howell2018}. The contrast curve for observations at 562 and 832 nm can be seen in Figure \ref{fig:contrast}, which shows the detection limits in contrast ($\Delta m$) versus angular separation from PSF center in arcseconds for each wavelength. The inset image is the speckle auto-correlation function for the observation at 832 nm.

We detect no companion within one arcsecond down to a $\Delta m > 4$ in optical and $\Delta m > 6$ in the near-infrared (NIR). Beyond 0.2 arcseconds, we see no spikes in the contrast curve above $\Delta m > 4$ in the 562 nm observation and $\Delta m > 5$ in the 832 nm observation, implying no bright, close companions to \hoststar. Difference image centroiding performed in the SPOC data validation tests for this system constrain the location of the transit source to within 1 arcsecond of \hoststar and thereby complement the high-resolution imaging results. Combined with the confirmed transit ephemeris and duration from the SG1 transit observation, we conclude that \hoststar is the source of the transit signal.

\subsection{Radial Velocities}

Radial velocity observations were obtained for this target using three instruments, and can be found in table \ref{tab:rvs}. Each radial velocity has been zero-subtracted using the best fit instrumental zero-point offset from the model, and the table is sorted in time. \hoststar was monitored with the fibre-fed FEROS spectrograph mounted on the MPG 2.2m \citep{Kaufer99} telescope at La Silla Observatory, in Chile. Thirteen spectra were obtained between January 3 and March 16 of 2020. These observations were performed in the context of the Warm gIaNts with tEss \citep[WINE,][]{brahm19,Schlecker2020} collaboration. We adopted an exposure time of 1200 s, and the observations were performed with the simultaneous wavelength calibration mode to trace the instrumental radial velocity drifts. The source in the secondary fiber was a Thorium-Argon lamp. FEROS data was processed with the \textsf{ceres} pipeline \citep{Brahm17CERES} which delivers precision radial velocities and bisector span measurements through cross-correlation with a G2-type binary mask. A future analysis with a template spectrum more similar to \hoststar may provide reduced radial velocity uncertainties. We found no significant correlation between the radial velocity and bisector span measurements. The signal-to-noise ratio per resolution element of these spectra ranged from 60 to 80. The spectral analysis routine included in \textsf{ceres} shows that \hoststar has a solar-like temperature ($T_\text{eff}=$ 5800 $\pm$ 100 K) and metallicity, a sub-solar surface gravity (log($g$) = 3.9 $\pm$ 0.1), and a moderate projected rotational velocity ($v$sin$i$ = 7.5 $\pm$ 0.5 km/s).

RV observations were also obtained using the High Resolution Spectrograph (HRS) on the South African Large Telescope (SALT) in Sutherland, South Africa. 7 radial velocity observations were obtained between December 20, 2019 and October 17, 2020. These observations were made by observing the target through an iodine cell, and were reduced using a modified version of the pipeline described in \citet{butler1996}. Due to the relative faintness of this target, multiple template spectra of the target were coadded together, resulting in a final template spectrum with a signal-to-noise ratio $>$ 50. Measurements and measurement uncertainties described here are determined using the blue component of the spectrum falling onto the blue CCD detector.

Finally, we obtained four observations with the CHIRON optical echelle spectrometer \citep{tokovinin2013} on the SMARTS 1.5m telescope at CTIO between February 17, 2020 and March 6, 2020. Data were obtained in slicer mode, which uses an image slicer and fiber bundle to yield $R\approx79,000$\ over the spectral range 410 nm to 880 nm. We extracted RVs by modeling the least-squares deconvolution spectral line profiles \citep{donati1997}.

\begin{figure*}[ht!]
    \centering
    \includegraphics[width=\textwidth]{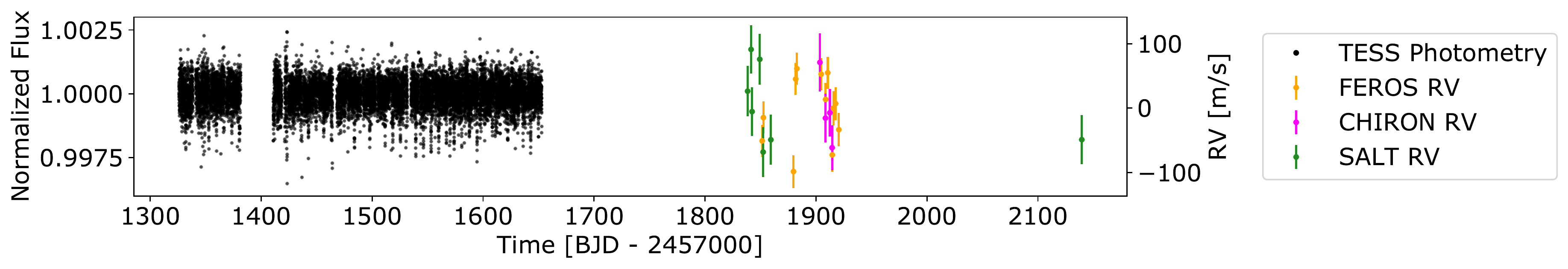}
    \caption{All time-series observations of \planet used in this analysis, including photometry and radial velocity measurements, plotted over time. The observations have independent y-axes: photometry plotted in normalized flux units on the left y-axis and radial velocity plotting in m s$^{-1}$ on the right y-axis.}
    \label{fig:full_phot_rv}
\end{figure*}

\begin{table}[]
    \centering
    \begin{tabular}{l c r}
        \hline
        Instrument & Time (JD - 2457000) & Relative RV (m/s)  \\
        \hline
        SALT & 1838.398 & 26.1 $\pm$ 15.8 \\
        SALT & 1841.495 & 91.0 $\pm$ 11.2 \\
        SALT & 1842.409 & -5.6 $\pm$ 12.2 \\
        SALT & 1849.349 & 75.7 $\pm$ 16.9 \\
        FEROS & 1851.541 & -50.7 $\pm$ 11.9 \\
        SALT & 1852.382 & -68.4 $\pm$ 15.2 \\
        FEROS & 1852.592 & -14.5 $\pm$ 11.9 \\
        SALT & 1859.345 & -49.4 $\pm$ 14.6 \\
        FEROS & 1879.763 & -98.4 $\pm$ 13.0 \\
        FEROS & 1881.709 & 45.3 $\pm$ 11.7 \\
        FEROS & 1882.812 & 61.6 $\pm$ 12.0 \\
        CHIRON & 1903.547 & 72.4 $\pm$ 44.2 \\
        FEROS & 1904.693 & 52.9 $\pm$ 12.0 \\
        CHIRON & 1908.556 & -14.1 $\pm$ 37.1 \\
        FEROS & 1908.655 & 13.7 $\pm$ 13.3 \\
        FEROS & 1910.610 & 55.1 $\pm$ 11.8 \\
        CHIRON & 1912.558 & -5.8 $\pm$ 35.8 \\
        CHIRON & 1914.529 & -60.0 $\pm$ 34.0 \\
        FEROS & 1914.667 & -72.5 $\pm$ 15.0 \\
        FEROS & 1915.694 & 0.2 $\pm$ 14.8 \\
        FEROS & 1917.692 & 7.2 $\pm$ 12.9 \\
        FEROS & 1920.592 & -33.3 $\pm$ 14.1 \\
        SALT & 2139.559 & -49.2 $\pm$ 13.3 \\
        \hline
    \end{tabular}
    \caption{Radial velocities measured for \hoststar by the FEROS, CHIRON, and SALT instruments. The RVs have been zero-point subtracted based on the best fitting orbital model and sorted in time.}
    \label{tab:rvs}
\end{table}

\section{Host Star Characterization} \label{sec:hoststar}

\subsection{Stellar Parameters}

We characterized the host star by first performing a spectroscopic analysis of a co-added FEROS spectrum, with a S/N per resolution element ranging from $60-80$. To derive atmospheric parameters we used ZASPE \citep{zaspe}, yielding $\tstar=5966 \pm 80$\,K, $\loggstar = 3.71 \pm 0.15$\,dex, $\fehstar=0.15 \pm 0.05$\,dex and $\vsini \approx 8.0$\,km/s. An independent spectrosocopic analysis using iSpec \citep{ispec1,ispec2} yielded consistent results, with $\tstar \approx$\,6020\,K, $\loggstar \approx 3.88$\,dex, $\fehstar \approx 0.14$\,dex and $\vsini \approx 5.7$\,km/s. We furthermore extracted atmospheric parameters from GALAH DR3 \citep{galahdr3}, yielding $\tstar=5811$\,K and  $\fehstar=0.17$ dex. Photometric estimates of the effective temperature were calculated using 2MASS $J-K$ color-\tstar\ relation from \cite{casagrande11} and the TESS Input Catalog \citep{stassun18}, yielding $\tstar=5936$\,K, and $\tstar=5720$\,K, respectively. For the final atmospheric parameters we adopted the self-consistent solution from ZASPE with uncertainties calculated by adding in quadrature the formal uncertainty with the standard deviation over all independent \tstar\ estimates. For \fehstar\ we used the same procedure, but instead adding a model-dependent error of 0.062\,dex in quadrature \citep{torres12}. The final values are $\tstar=5966 \pm 136$\,K and $\fehstar = 0.14 \pm 0.08$\,dex.

To calculate additional parameters we combined \tstar\ and \fehstar\ with the Gaia DR2 parallax \citep{lindegren18} and 2MASS photometry \citep{skrutskie06} using the open-source code \textsf{isoclassify} \citep{huber17,berger20}. Specifically, we first used the ``grid-mode'' of \textsf{isoclassify} with \tstar, \fehstar, parallax and $K$-band magnitude to calculate a \loggstar\ value, which was then used in the ``direct-mode'' to interpolate a bolometric correction and calculate an isochrone-independent luminosity. Finally, \tstar, \fehstar\ and luminosity were again used as an input to the ``grid-mode'' to calculate estimates of stellar mass, density and age. We followed \cite{tayar2020} to calculate systematic errors due to different model grids, which were added in quadrature to our estimates for mass, density and age. The full list of stellar parameters is given in Table \ref{table:stellar}.

\begin{table}
\begin{center}
    \begin{tabular}{l  r}
        \hline
        \rule{0pt}{3ex}\textit{Target IDs} & \\
        \rule{0pt}{3ex}TOI & 2184 \\
        TIC & 176956893 \\
        TYC & 8907-998-1 \\
        2MASS & J06431993-6656515 \\
        Gaia DR2 & 5280444557068991616 \\
        \hline
        \rule{0pt}{3ex}\textit{Coordinates} & \\
        \rule{0pt}{3ex}RA & 6:43:20 \\
        Dec & -66:56:52 \\
        \hline
        \rule{0pt}{3ex}\textit{Characteristics} & \\
        \rule{0pt}{3ex}\tess magnitude & 11.4 \\
        $V$ magnitude & 12.3 \\
        $K$ magnitude & 10.4 \\
        Radius $R_\star$ $(R_\odot)$ & \starradius \\
        Mass $M_\star$ $(M_\odot)$ & \starmass \\
        Effective temperature $T_{\rm eff}$ (K) $ $ & \teff \\
        Surface gravity $\log(g)$ (dex) & \logg \\
        Iron abundance $ $[Fe/H] (dex) $ $ & \feonh \\
        Age (Gyr) $ $ & \age \\
        Density $\rho_\star$ (g cm$^{-3}$) & \starrhogcm \\
        \vsini (km/s) & $7.5 \pm 0.5$ \\
        \hline
   \end{tabular}
	 \caption{Host star properties.} 
	 \label{table:stellar}
\end{center}
\end{table}

\begin{figure}[ht!]
    \centering
    \includegraphics[width=.45\textwidth]{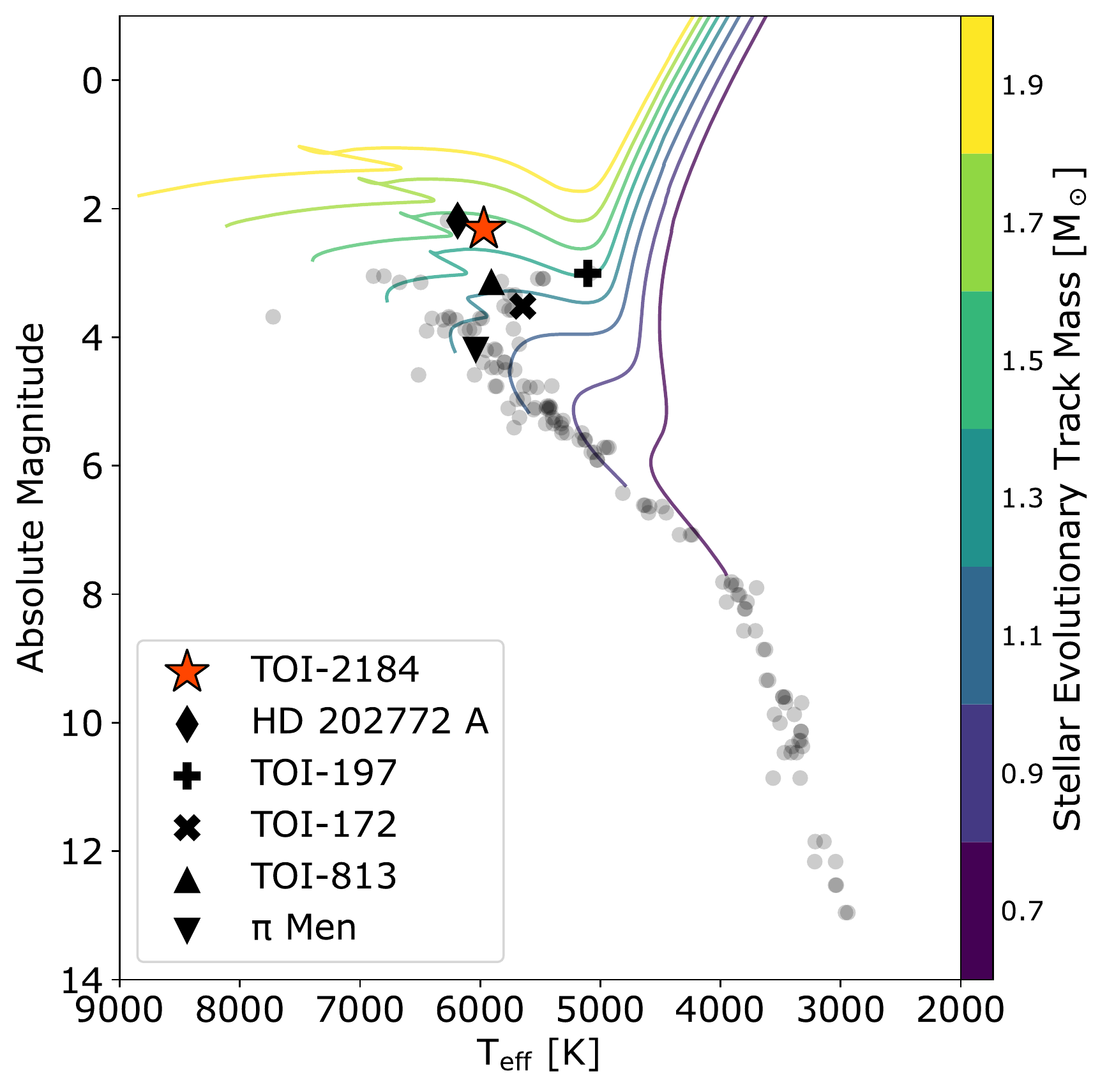}
    \caption{H-R diagram of confirmed exoplanet host stars from \tess. The position of \hoststar is marked by the red star. Gray circles show the positions of confirmed \tess planet host stars. We also mark the positions of similar or otherwise notable planet host stars. These values were downloaded from the NASA Exoplanet Archive.}
    \label{fig:hr}
\end{figure}

The temperature, radius and mass of \hoststar (\teff K, \starradius $R_\odot$, \starmass $M_\odot$) demonstrate that the host star is an intermediate-mass subgiant which is currently evolving towards the red-giant branch. Figure \ref{fig:hr} shows the position of \hoststar\ on a Hertzsprung-Russell (H-R) diagram, using tracks from the MESA Isochrones \& Stellar Tracks (MIST; \citealt{dotter2016}; \citealt{choi2016}; \citealt{paxton2011}). Because stars move through the subgiant phase relatively quickly, few planet hosts have been discovered in this regime and a number of evolutionary processes of planets orbiting subgiants are poorly understood. Indeed, \hoststar\ is one of the most massive stars with a transiting planet detected by \tess, and occupies a region of the H-R diagram with few detections. Better constraints on stellar parameters may be possible with asteroseismic analysis of \tess extended mission data, given that oscillation amplitudes are large enough to be detected.



\subsection{Stellar Rotation}


A search for stellar variability was conducted using the Causal Pixel Modeling (CPM; \citealt{wang2016,hattori2021}) approach adapted for \tess\footnote{\href{https://github.com/soichiro-hattori/unpopular}{github.com/soichiro-hattori/unpopular}}. This analysis returned tentative trends of $\sim$155 days and $\sim$73 days. However, long-period variability is challenging to measure with \tess due to the relatively short observing windows, and the CPM method remains fairly untested for \tess data. We conclude that these results should be interpreted with caution, and we report no definitive rotation period recovered from \tess photometry for \hoststar.

An additional analysis of stellar variability using data collected by the Wide Angle Search for Planets (WASP; \citealt{pollacco2006}) South found no rotational modulation for this target in the range from 2 to 100 days. This was conducted using 24,000 data points from four consecutive years, covering a span of about 160 nights each year. The upper limit of photometric variability detection for this target made by WASP-South is roughly 0.8 millimagnitudes. 

\section{Planet Characterization} \label{sec:planet}

We used the \exoplanet Python package \citep{exoplanet:exoplanet} to simultaneously fit an orbital model to the photometry and radial velocity observations. The data input to our model were the 23 radial velocity observations and 11 sectors of \tess 30-minute cadence photometry (Figure \ref{fig:full_phot_rv}).

Our \exoplanet model was constructed using the built-in solution for Kepler's equation. For limb darkening, we used a two-parameter quadratic model with normal distributions with mean values selected as the nearest grid point in Table 25 of \cite{claret2017}, which reports pre-computed quadratic limb darkening coefficients specifically for \tess for a variety of $T_{\rm eff}$, [Fe/H], and $\log{(g)}$. We selected our stellar parameters based on the best fit derived from \textsf{isoclassify}. We parameterized eccentricity using the \cite{kipping2013a} Beta distribution, which we favored over the \cite{vaneylen2019} distribution as the latter was derived for small planets. The other transit parameters we optimized were radius ratio $R_P/R_\bigstar$, impact parameter $b$, orbital period $P$, and midtransit time at a reference epoch $t_0$. The radial velocity components were parameterized with a separate RV offset and jitter term for each of the three instruments. To estimate mass, we optimized the semi-amplitude $K$ of the RVs. Our prior distributions can be found in Table \ref{table:planet}.

These distributions were created within a \textsf{PyMC3} model \citep{exoplanet:pymc3}, allowing us to optimize the model parameters using gradient descent. We sampled our optimized model parameters using No U-Turn Sampling (NUTS; \citealt{nuts}) with two chains of 4,000 draws, with 4,000 iterations used to tune the model. We determined the median and standard deviation for each of our model parameters from the sampled posterior distributions. To ensure that our chains converged, we checked the Gelman-Rubin $\hat{R}$ statistic \citep{gelman1992} and measured a value less than 1.007 for all model parameters.

\begin{figure}[ht!]
    \centering
    \includegraphics[width=.45\textwidth]{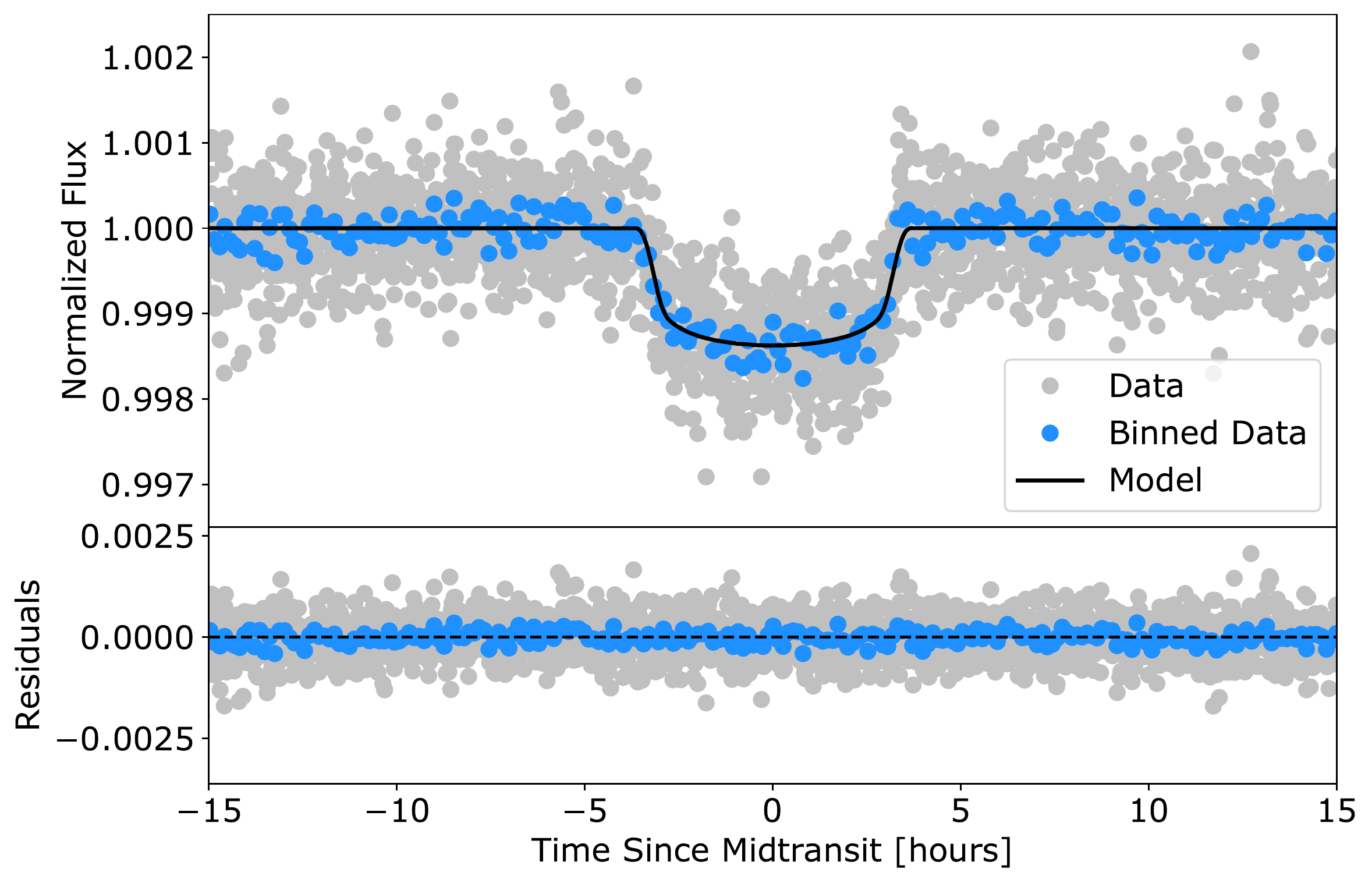}
    \caption{Light curve of \hoststar folded at a period of \period days. The de-trended photometry is shown in gray with the binned photometry overplotted in blue. The fit transit model is the solid black line in the top panel, and the bottom panel shows the residuals between the light curve and transit model. The slightly positive out-of-transit values in the residuals are a remnant of the trend seen in Figure \ref{fig:evenodd_compare}.}
    \label{fig:transit}
\end{figure}

\begin{figure}[ht!]
    \centering
    \includegraphics[width=.45\textwidth]{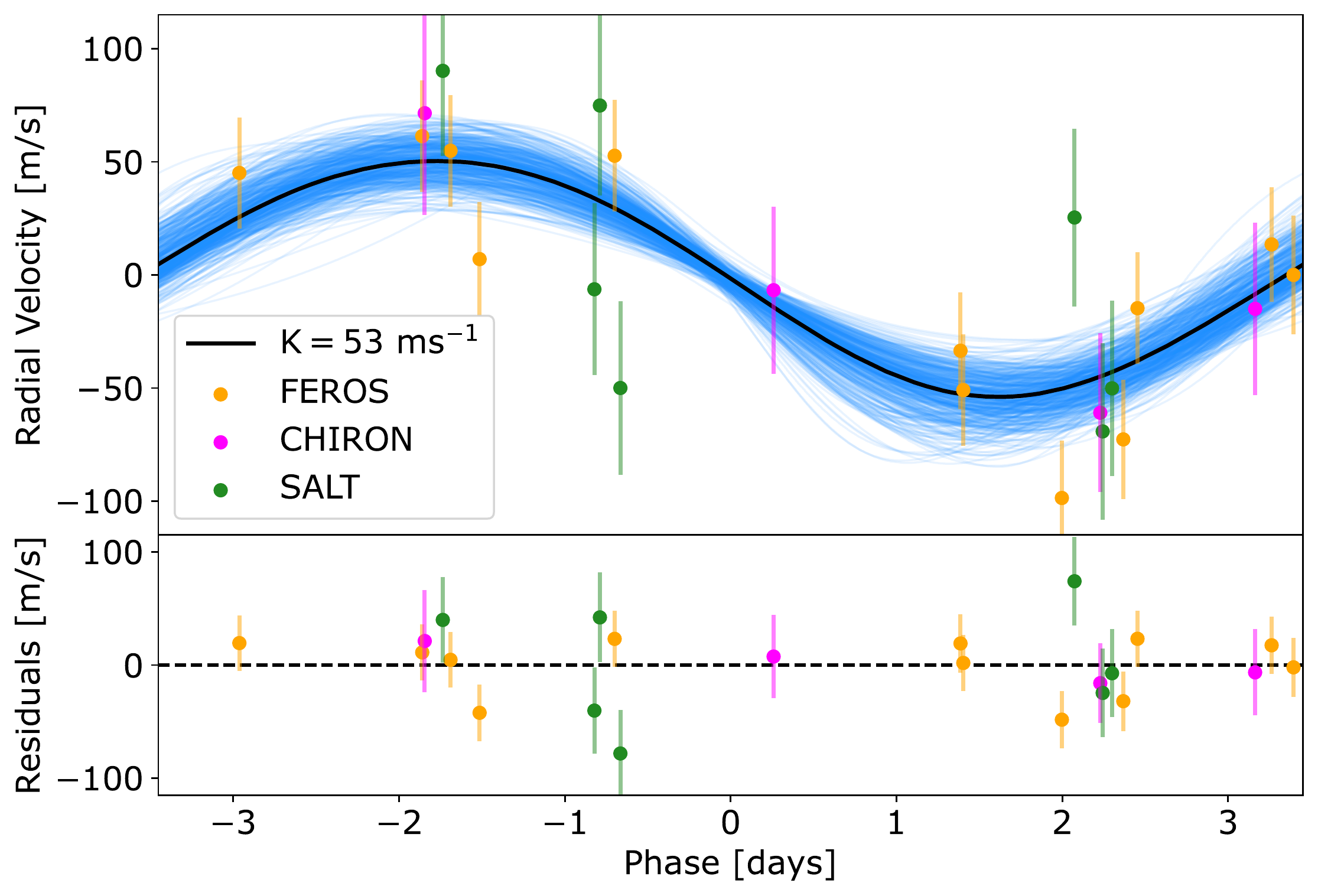}
    \caption{Phase-folded radial velocity measurements of \hoststar. The bottom panel is the residuals after subtracting the median radial velocity model from the data. The solid black line shows the best fit to the radial velocity. The blue lines each represent a single realization of the model drawn from the sampled posterior distribution, with 500 samples shown in total.}
    \label{fig:rv_fold}
\end{figure}

A full table of the parameters used in our model and their inferred values can be found in Table \ref{table:planet}. While the fit was performed simultaneously, the individual transit and RV components of the fit can be found in Figure \ref{fig:transit} and Figure \ref{fig:rv_fold}, respectively. The posterior distributions for a number of key model parameters are shown in the corner plot in Figure \ref{fig:corner} in the Appendix.

\begin{table*}
\begin{center}
    \begin{tabular}{l l r}
        \hline 
        Parameter & Prior & Value \\
        \hline
        \rule{0pt}{3ex}\textit{Fit Parameters} & & \\
        \rule{0pt}{3ex}$R_P/R_\bigstar$ & $\log\mathcal{N}[0.035, 0.03]$ & $0.0342 \pm 0.0007$ \\
        RV semi-amplitude $K$ (m s$^{-1}$) & $\log\mathcal{N}[55, 10] $ & \K \\
        Orbital period $P_{\text{orb}}$ (days) & $\log\mathcal{N}[6.9068, 0.1]$ &  \period \\
        Transit epoch $t_0$ (BJD - 245700) & $\mathcal{N}[1332.148, 0.1]$ & \transittime \\
        Transit duration $T_{\rm dur}$ (hours) & $\mathcal{N}[6.5, 1.0]$ & $6.78 \pm 0.07$ \\
        Impact parameter $b$ & $\mathcal{U}[0, 1+R_P/R_\bigstar]$ & $0.67\pm0.10$\\
        Eccentricity $e$ & $P_\beta(e\in[0,1])^\text{(a)}$ & \ecc\\
        Limb-darkening coefficient $q_1$ & $\mathcal{N}[0.2577, 0.1]^\text{(b)}$ & $0.2581 \pm 0.0097$ \\
        Limb-darkening coefficient $q_2$ & $\mathcal{N}[0.3034, 0.1]^\text{(b)}$ & $0.303 \pm 0.010$ \\
        \hline
        \rule{0pt}{3ex}\textit{Derived Physical Parameters} & & \\
        \rule{0pt}{3ex}Planet radius $R_p$ ($R_J$) & & \planetradius \\
        Planet mass $M_p$ ($M_J$) & & \planetmass \\
        Planet density $\rho_P$ (g cm$^{-3}$) & & \planetrhocgs \\
        Incident flux $F$ ($F_\oplus)$ & & $1429\pm151$ \\
        \hline
   \end{tabular}
	 \caption{Parameters for \planet. $\mathcal{N}[\mu,\sigma]$ denotes a normal (Gaussian) distribution with mean $\mu$ and standard deviation $\sigma$, and $\mathcal{U}[a,b]$ denotes a uniform distribution from $a$ to $b$. \textit{Notes:} $^\text{(a)}$This parameterization is described by the Beta distribution in \cite{kipping2013a}. $^\text{(b)}$Prior values retrieved from Table 25 of pre-computed limb-darkening coefficients for \tess by \cite{claret2017}.}
	 \label{table:planet}
\end{center}
\end{table*}

We measure the radius of \planet to be \planetradius $R_J$. The semi-amplitude derived from our best fit to the RV observations was \K m s$^{-1}$, from which we estimate a mass of \planetmass $M_J$. These can be used to estimate the density $\rho_P$, which we calculate to be $\rho_P=$ \planetrho $\rho_J$ (\planetrhocgs g cm$^{-3}$).

\section{Discussion} \label{sec:discussion}



\subsection{Radius Inflation} \label{sec:inflation}


A persistent mystery about hot Jupiters is the observed distribution of anomalously large planetary radii. Studies of planet inflation invoke different atmospheric processes, such as Ohmic heating driven by interactions between a planet's magnetic field and ionized winds in its atmosphere, to transfer energy from the surface to the planetary interior (e.g. \citealt{batygin10, thorngren2018, komacek2020, thorngren2021}). These models are sensitive to a number of observable properties, such as star and planet mass and composition, metallicity, and orbital period. 

\begin{figure}[ht!]
    \centering
    \includegraphics[width=.45\textwidth]{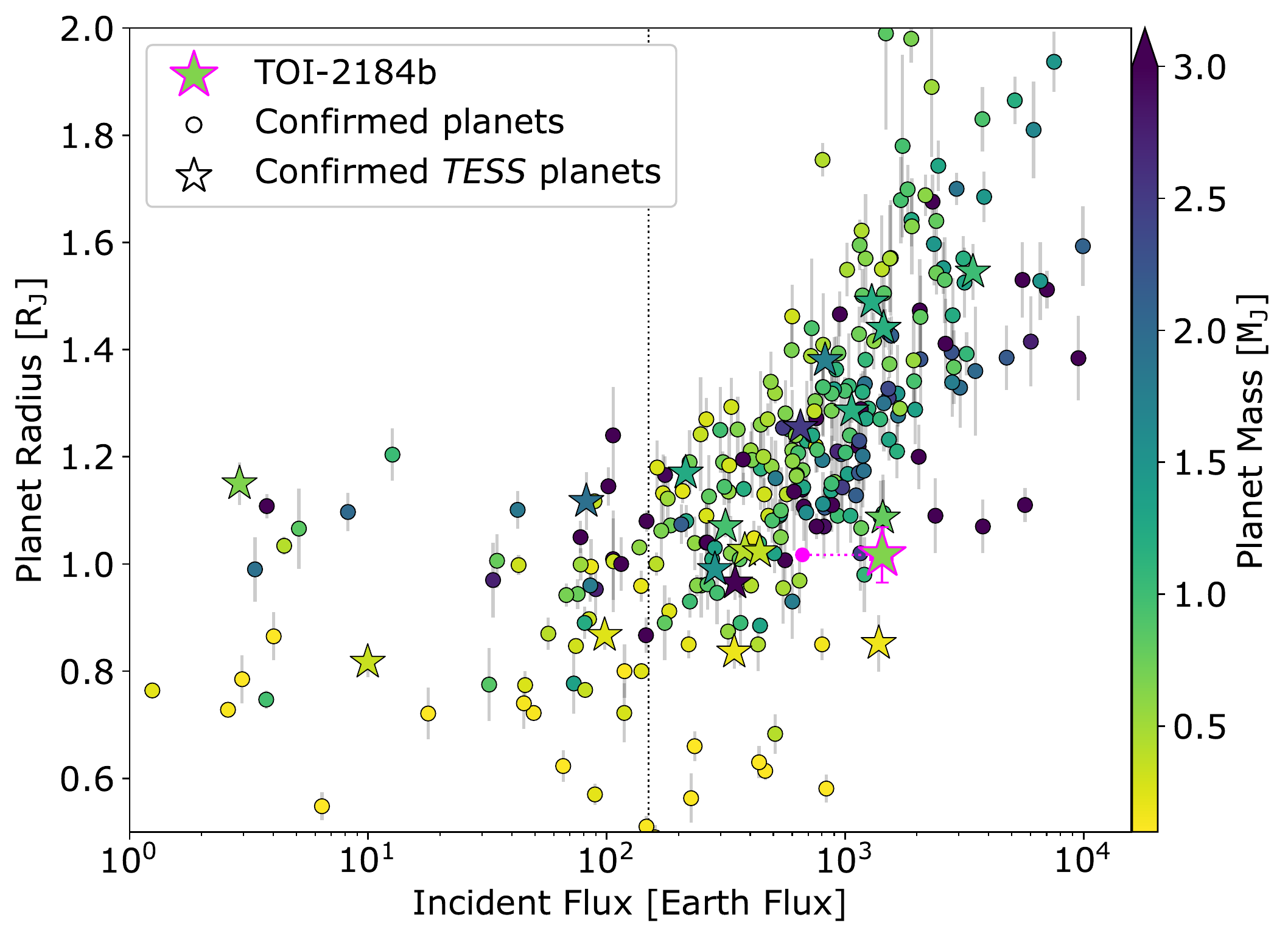}
    \caption{Incident flux received by the planet versus planet radius. Planets discovered by \tess are marked by stars, while those discovered by different instruments are marked by points. \planet is the star outlined in pink, with the estimated main sequence position shown by the pink point connected with a dashed line. The vertical dashed line shows the \cite{demory2011} threshold for inflation.}
    \label{fig:rad_v_flux}
\end{figure}

To place our planet in context with hot Jupiter inflation, Figure \ref{fig:rad_v_flux} shows the radii of confirmed Jovian exoplanets versus the intensity of incident flux they receive. For clarity, we made cuts which limit the sample to planets with mass precision $<30\%$ and radius precision $<10\%$. There is a strong correlation between incident flux and radius, with radius increasing as a planet receives higher incident flux. 

Using a sample from \kepler, \cite{demory2011} find a lower limit for incident flux before measurable inflation occurs among hot Jupiters to be $\sim$2$\times 10^8$ erg s$^{-1}$ cm$^{-2}$ ($\sim$150 $F_\oplus$). We calculate the incident flux received by \planet to be $1429\pm151$ $F_\oplus$, well above this nominal lower limit for inflation. While the limit established by \cite{demory2011} is not a hard cutoff, it places \planet within a regime where some degree of inflation is commonly observed and theoretically possible. Despite this level of incident flux, \planet shows no definitive evidence for significant inflation, and has a density much closer to that of Jupiter relative to other confirmed hot Jupiters of similar mass and incident flux.

\begin{figure}[ht!]
    \centering
    \includegraphics[width=.45\textwidth]{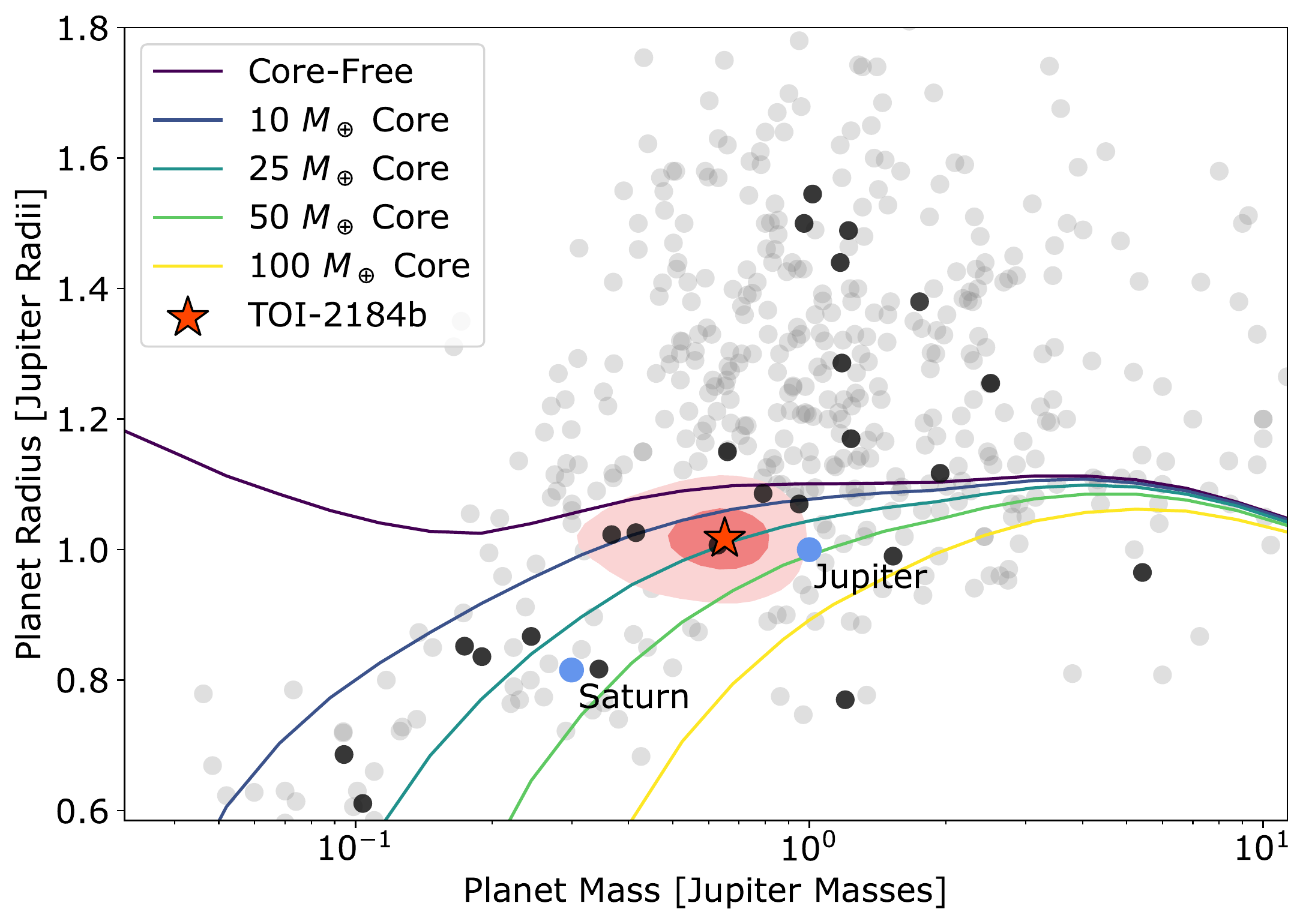}
    \caption{Planet mass versus planet radius for confirmed Jovian exoplanets. Confirmed planets discovered by \tess are shown as black points while planets discovered by other telescopes are shown in gray. The colored lines show models by \cite{freedman2014}. These models do not account for planetary inflation, which explains the large radii of the many planets which fall above these curves. \planet is marked by the red star, and the shaded regions represent the 1- and 2-$\sigma$ confidence intervals.}
    \label{fig:mass_radius_hj}
\end{figure}

Theories for inflation are strongly dependent on planet mass, so it is important to consider the relationship between the radius of \planet and that of other similar mass hot Jupiters. In Figure \ref{fig:mass_radius_hj}, we plot the planet radius versus planet mass for all confirmed Jovian exoplanets. When compared to other confirmed planets with masses within 0.1 $M_J$ of \planet, this system falls in the 12$^{\rm th}$ percentile in radius. When compared to other planets within the same mass range which have incident flux reported as $F > 150$ $F_\oplus$, \planet has the smallest radius\footnote{Data retrieved from the \href{https://exoplanetarchive.ipac.caltech.edu/}{NExScI archive} on March 17, 2021.}. Based on the distribution of similar Jupiter-mass planets, \planet falls below the general trend of increased radius with increased incident flux.


Using the \cite{weiss2013} relationship for predicted radius (for $M_P > 150$ $M_\oplus$)
\[
\frac{R_P}{R_\oplus} = 2.5 \left( \frac{M_P}{M_\oplus} \right)^{-0.039} \left(\frac{F}{\text{erg s}^{-1}\text{cm}^{-2}} \right)^{0.094}
\]
we predict a planetary radius of $\sim 14.5$ $R_\oplus$, more than 5-$\sigma$ greater than the measured radius of $\sim 11.4$ $R_\oplus$. Relative to other similar planets and based on the current intensity of incident flux, the small radius of \planet makes it an outlier from the expected inflation trend. In Figure \ref{fig:rad_v_flux}, the predicted radius places \planet much closer to the observed radii of inflated hot Jupiters. Because the timescale and causes of inflation are uncertain, the apparent lack of inflation in this system can provide useful insights into the process.


One theoretical explanation for the inflation of hot Jupiters is delayed radiative cooling and contraction \citep{lopez2016}. In this scenario, Jovian planets on close-in orbits receive a level of incident flux which prohibits their contraction while their host star is on the main sequence. Another possible scenario is re-inflation, in which the planet cools and contracts while on the main sequence and then, as the host star evolves off of the main sequence, crosses a threshold for incident flux which causes the planetary atmosphere to expand. 


An important feature of this system to consider is how the incident flux and radius of \planet today compare to when its host star was on the main sequence. This will help to determine how the evolution of \planet fits into the timeline of re-inflation models. The predicted main sequence incident flux of \planet is indicated by the pink point in Figure \ref{fig:rad_v_flux}, connected to its current position by a dashed line. This value was calculated by estimating the luminosity of \hoststar from a standard mass-luminosity relationship and recalculating the incident flux using the estimated main sequence luminosity, resulting in an estimated zero-age main sequence incident flux of $\sim 660$ $F_\oplus$.

The detection of a transiting planet during its host star's ascent of the giant branch is particularly valuable because the point at which planetary atmospheric inflation begins is still unclear. Re-inflation is expected to occur when the incident flux received by a planet exceeds $\sim150$ $F_\oplus$ \citep{lopez2016,demory2011}. Supporting evidence for this theory has been found in two planets which have lower equilibrium temperatures but orbit more evolved stars when compared to \planet \citep{grunblatt16, grunblatt2017, jones2018}. When compared to these systems with observed inflated hot Jupiters, \planet receives a higher incident flux (in excess of $1400$ $F_\oplus$) and its host star is much less evolved ($<3.5$ $R_\odot$). 
As its host star continues to evolve onto the RGB, \planet will continue to receive more intense irradiance from its increasingly luminous host.  However, the main sequence incident flux likely already exceeded the nominal inflation threshold. Given that the flux incident on this planet has always been above the threshold for inflation, it is unlikely that the planet's lack of inflation is caused by the host star's early stage of evolution onto the giant branch. This can be seen in Figure \ref{fig:rad_v_flux}, where both the current and estimated main sequence position of \planet fall above $\sim150$ $F_\oplus$.  


We also consider whether the planet radius may be underestimated through the systematics discussed in \S \ref{sec:evenodd}. However the average difference in depth between transits taken within a day of a data gap and all other transits is only on the order of 20\%, and these near-gap transits only constitute 10 of the 38 full transits observed in sectors 1-12. This makes it unlikely that this effect creates a discrepancy to the degree implied by our predicted radius. Another potential non-astrophysical source of radius anomaly is uncertainty in measured transit depth from our light curve generation methods. Our photometry pipeline does not account for ``blending" in the flux time series due to contamination by nearby stars, however the MIT QLP applies a correction for the expected flux contribution by nearby stars based on their \tess band magnitude \citep{huang2020}. We compared the depth of the QLP and \giants light curves, and found no significant depth difference ($\lessapprox5\%$) between the pipelines. Additionally, the light curves for \hoststar generated by the \eleanor pipeline produce consistent transit results. Similarly, the transit depth for this planet obtained by the SPOC Data Validation report of the \tess Extended Mission data agrees with our reported radius within errors. We conclude that the difference between the expected and measured radius is likely not caused by any systematic effects.


The lack of observed radius inflation of \planet suggests that it is a potential example of a planet caught in the early stages of re-inflation. However, the estimated high incident flux on the main sequence suggests the explanation for the lack of inflation could be unrelated to stellar evolution. A larger sample of hot Jupiters will be required to establish a timeline for planetary inflation and its relationship with host star evolution.


\subsection{Eccentricity}

A suggested formation pathway for hot Jupiters is that they arrive in their current position by migrating from long-period highly eccentric orbits to shorter-period circularized orbits \citep{dawson2018}. Despite the small sample of confirmed planets around evolved stars, interesting trends in this population have been identified. What remains unclear is when this happens, and the population of planets around evolved stars has yielded interesting insights into this question. Specifically, \cite{grunblatt2018} found that giant planets ($R_p > 0.4R_J$) orbiting evolved stars on short ($<30$ day) orbital periods tend to have significantly higher eccentricity than giant planets orbiting dwarfs. This trend may originate from the changes in tidal migration caused by stellar evolution---when stars evolve into subgiants, their radii increase, causing them to be more strongly affected by tidal effects and accelerating tidal migration. In this scenario, close-in planets, whose orbits have had time to tidally circularize nearer to the host star, will be consumed by the star's growing radius, exacerbated by the orbital decay caused by more rapid tidal dissipation, while longer period planets will migrate into a nearer orbit while still maintaining a modest eccentricity (\citealt{villaver2009, villaver2014}).


Our model finds a low eccentricity for \planet of \ecc. In most respects, \planet is similar to the sample in \cite{grunblatt2018}---which analyzes the eccentricities of close-in giant planets discovered by \kepler---with a period of 6.9 days, Jupiter-like mass and radius, and host star radius of 2.9 $R_\odot$. The study by \cite{grunblatt2018} finds that, for \kepler, close-in giant planets orbiting evolved hosts have a median eccentricity of $e\approx0.152$ compared to $e\approx0.056$ for close-in giant planets orbiting dwarfs. The deviation from this trend by \planet is potentially explained by the apparent earlier evolutionary stage of \hoststar. Further photometric and radial velocity observations could distinguish whether this planet is more similar to the main sequence or evolved population, or if it occupies a ``transition zone" between the two. Based on the earlier evolutionary stage of \hoststar, the planet likely migrated well before the star evolved off of the main sequence, and thus fully circularized. 

More precise measurements of the eccentricity of \planet through RVs may help clarify the migration scenario, however the stellar jitter associated with subgiants likely limits the achievable measurement precision. Considering the population more broadly, future detections could fall into two categories---if a similarly non-inflated planet is observed with low eccentricity to high precision, the relationship between eccentricity and inflation will remain unclear. However, if a non-inflated planet is detected with high eccentricity, it would contradict mechanisms for tidal inflation and requires additional pathways to inflated hot Jupiters. Additionally, a greater volume of detections would help to discern whether most hot Jupiters stay relatively stable and circular around evolved stars to large ages, or if we are only seeing the remnants of once much cooler planets on eccentric orbits around main sequence stars.

\section{Conclusions} \label{sec:conclusions}

We have begun a search for planets around evolved host stars using the \tess Full Frame Image data. Our search yielded the discovery of \planet, a hot Jupiter around a massive subgiant that was initially discarded as a false positive by the QLP pipeline.

Our main conclusions are as follows:

\begin{itemize}
    \item The coincidental relationship between the orbital period of \planet and the orbital period of the \tess spacecraft ($P_{\text{orb, } TESS} \approx 2P_{\text{orb, planet}}$) caused a systematic difference in measured depth of alternating transits. This caused the transit signal of \planet to mimic the primary/secondary eclipse signal expected from eclipsing binary systems, and the candidate was rejected by planet discovery pipelines. The misclassification of \planet implies that a number of \tess planets near harmonics of the spacecraft's orbital period may have been missed or misclassified as false positives.
    \item We used TESS photometry and ground-based radial velocities to find a radius of \planetradius $R_J$ and a mass of \planetmass $M_J$. We estimate the incident flux received by planet to be $1429\pm151$ $F_\oplus$. Compared to other planets of similar mass and incident flux, \planet is among the smallest hot Jupiters, with no evidence of major atmospheric inflation. This detection occupies a poorly understood phase of the post-main-sequence evolution of planetary systems and provides clues to the physical mechanism(s) behind the radius inflation of hot Jupiters.
    \item Compared to other planets in a similar regime, \planet exhibits low eccentricity. More precise measurement of this system's eccentricity will determine whether it follows previously suggested trends of higher-eccentricity planets around evolved hosts, and will help place constraints on the timescales of tidal inspiral and eccentricity decay as host stars evolve up the red giant branch.
\end{itemize}

After being designated as a TOI, this system has received 2-minute cadence observations from the \tess extended mission. With another year of photometry from \tess, future analysis may place tighter constraints on transit parameters of \planet. These short-cadence observations may also probe the oscillations of its subgiant host. 

\planet is the first discovery of our survey, and several additional detections of planets around evolved stars are forthcoming (Grunblatt et al., in prep). In addition to generating our light curves and summary plots, the \giants pipeline stores output values from the BLS search. These will be used in future work, along with the light curves, to perform an automated search of the data. As \tess continues to observe, the observation baseline for potential targets increases allowing for more precise characterization of planet transit parameters. The higher cadence FFI observations taken in the \tess extended mission will also improve transit parameter precision as well as open the door for more detections of stellar oscillations. The light curves produced by our \giants pipelines are also being used for stellar astrophysics and other applications, for example a study of asteroseismic detections in the \kepler field by \tess \citep{stello2021}. The new cadence will push the Nyquist frequency for FFI targets higher and allow asteroseismic characterization of an increased sample of less evolved host stars. 

\planet exemplifies a particularly unlucky case of systematic trends confounding planet detection methods, but still shows the ways in which periodic instrumental trends can produce false negatives. Its detection demonstrate why a more focused search for planets around evolved targets is warranted. There remains a wealth of planets orbiting faint stars ($T$ mag $>12$) in the \tess FFIs, and by targeting evolved stars, our search will produce a statistical sample of planets which can be used to test the connection between stellar evolution and planet demographics.

\section*{Acknowledgements}

This work relied heavily on open source software tools, and we would like to thank the developers for their contributions to the astronomy community. We would also like to thank Chelsea Huang for providing clarifying insight into the MIT Quick-Look Pipeline (QLP) vetting process. N.S., S.G.\ and D.H.\ acknowledge support by the National Aeronautics and Space Administration under Grant 80NSSC19K0593 issued through the \tess Guest Investigator Program. N.S.\ and A.C.\ acknowledge support from the National Science Foundation through the Graduate Research Fellowship Program under Grant 1842402. Any opinions, findings, and conclusions or recommendations expressed in this material are those of the authors and do not necessarily reflect the views of the National Science Foundation. D.H.\ also acknowledges support from the Alfred P. Sloan Foundation. A.J.\ and R.B.\ acknowledge support from FONDECYT project 1210718, and ANID - Millennium Science Initiative - ICN12$\_$009. This work makes use of observations from the LCOGT network. Part of the LCOGT telescope time was granted by NOIRLab through the Mid-Scale Innovations Program (MSIP). MSIP is funded by NSF. We acknowledge the use of public \tess data from pipelines at the TESS Science Office and at the \tess Science Processing Operations Center. Resources supporting this work were provided by the NASA High-End Computing (HEC) Program through the NASA Advanced Supercomputing (NAS) Division at Ames Research Center for the production of the SPOC data products.

We extend deep gratitude to those of Hawaiian ancestry, upon whose Mauna we are lucky to be guests. 

\facilities{CTIO, LCOGT, SALT, TESS}

\software{\astroimagej \citep{Collins:2017}, \astropy \citep{astropy2013,astropy2018}, \eleanor \citep{feinstein2019}, \exoplanet \citep{exoplanet:exoplanet} and its dependencies \citep{exoplanet:agol19, exoplanet:exoplanet, kipping2013b, exoplanet:luger18, exoplanet:pymc3, exoplanet:theano}, \lightkurve \citep{lightkurve}, \textsf{matplotlib} \citep{matplotlib}, \textsf{numpy} \citep{numpy1, numpy2}, \textsf{scipy} \cite{scipy}, \tapir \citep{Jensen:2013}, \tesscut \citep{brasseur2019}.}

\pagebreak
\nocite{tange2018}
\bibliography{references,references2}{}
\bibliographystyle{aasjournal}

\appendix \label{appendix}

\begin{figure}[ht!]
    \centering
    \includegraphics[width=.75\textwidth]{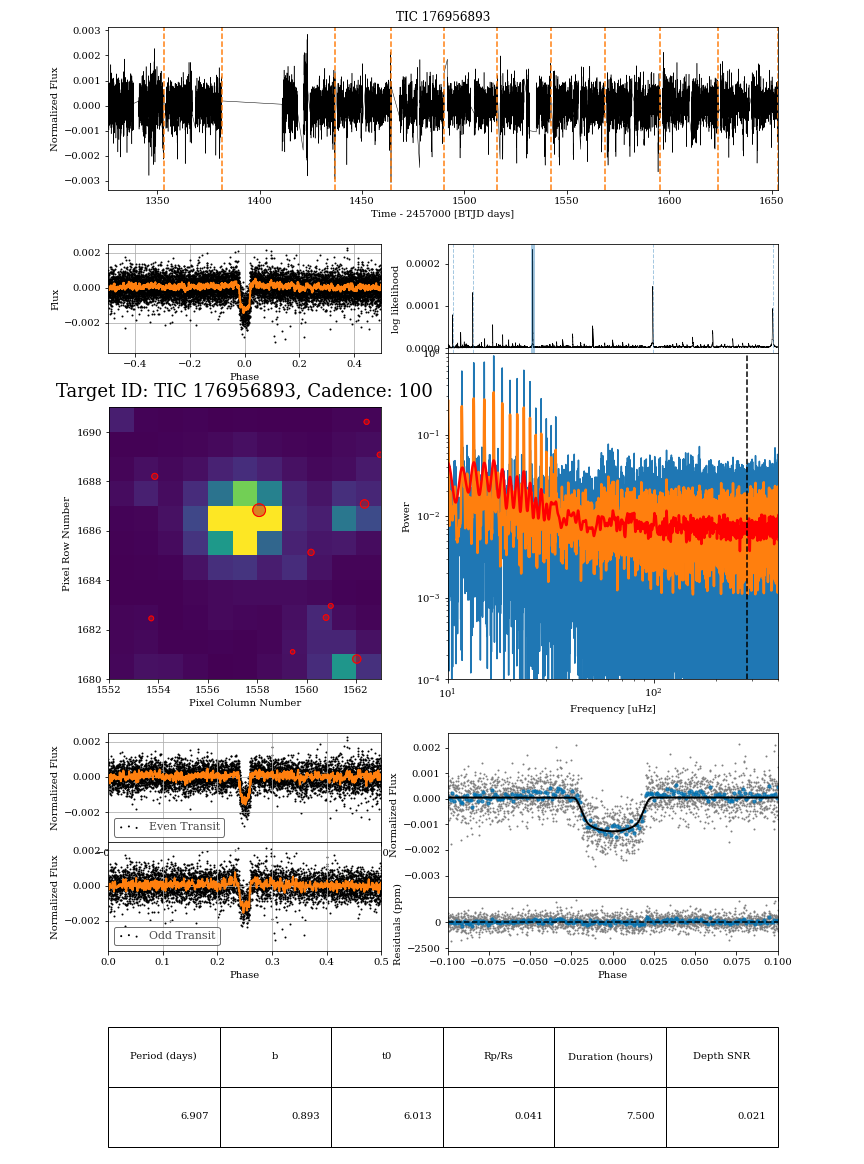}
    \caption{One-page vetting summary for \planet used to identify the transit signal.}
    \label{fig:vetting}
\end{figure}

The summary plot that was used in the discovery of \planet is shown in Figure \ref{fig:vetting}. This one-page pdf summary contains the elements described in \ref{sec:search}, including the full de-trended light curve, Box-Least Squares and Lomb-Scargle periodograms, and preliminary transit model fit. The target pixel file cutout image (center left) includes an overlay of nearby Gaia sources (red circles) to help rule out background contaminants, and shows no nearby bright neighbors that could significantly confound the detection of \planet. The estimated transit parameters from the initial fit are reported in the table at the bottom of the summary.

\begin{figure}[ht!]
    \centering
    \includegraphics[width=.75\textwidth]{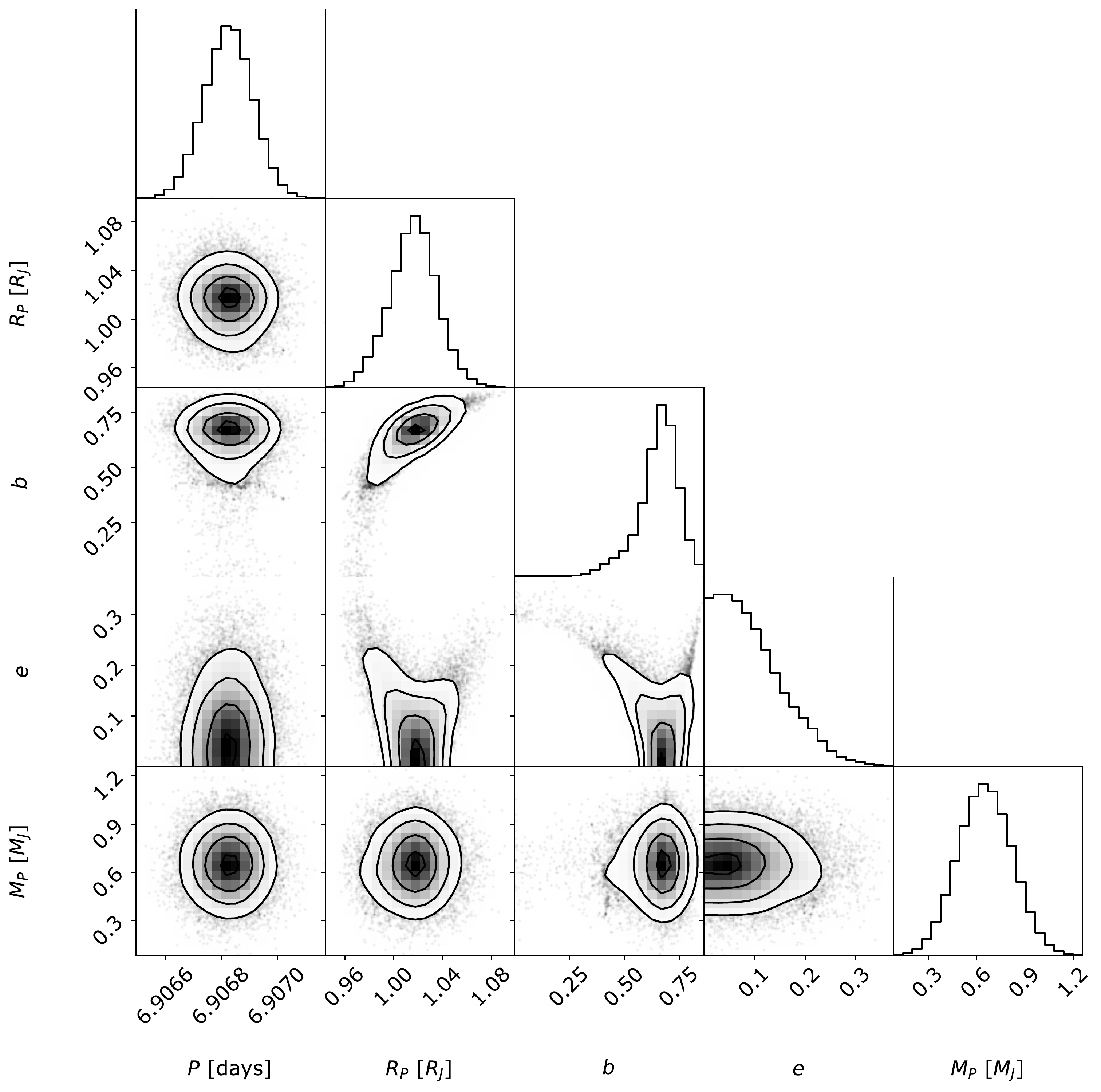}
    \caption{Posterior distributions and correlations between parameters in our orbital model.}
    \label{fig:corner}
\end{figure}

We also include a corner plot of the posterior distributions and correlations between parameters in our optimized and sampled orbital model in Figure \ref{fig:corner}.

\end{document}